\documentclass[]{elsarticle}

\usepackage{lineno,hyperref}
\usepackage{float}

\usepackage{subcaption}
\usepackage{color}
\usepackage{latexsym}

\journal{Journal of \LaTeX\ Templates}

\usepackage{amsmath}
\usepackage{amssymb}

\usepackage[usenames, dvipsnames]{xcolor}
\usepackage{tikz}

\begin{document}

\newcommand{\rectangle}{\raisebox{0pt}{\tikz{\draw[gray,solid,line width = 1.0pt](2.mm,0) rectangle (3.5mm,1.5mm);\draw[-,gray,solid,line width = 1.0pt](0.,0.8mm) -- (5.5mm,0.8mm)}}}

\newcommand{\trianglee}{\raisebox{0pt}{\tikz{\draw[gray,solid,line width = 1.0pt](4.05mm,0.2mm) -- (1.55mm,0.2mm) -- (2.75mm,1.65mm) -- cycle;\draw[-,gray,solid,line width = 1.0pt](0.,0.8mm) -- (5.5mm,0.8mm)}}}

\newcommand{\cross}{\raisebox{0pt}{\tikz{\draw[-,gray,solid,line width = 1.0pt](2.75mm,0.0mm) -- (2.75mm,1.6mm);\draw[-,gray,solid,line width = 1.0pt](0.,0.8mm) -- (5.5mm,0.8mm)}}}

\newcommand{\circlee}{\raisebox{0pt}{\tikz{\draw[-,gray,solid,line width = 1.0pt,fill=gray] (2.75mm,0.8mm) circle (0.75mm);\draw[-,gray,solid,line width = 1.0pt](0.,0.8mm) -- (5.5mm,0.8mm)}}}

\newcommand{\redline}{\raisebox{0pt}{\tikz{\draw[-,red,solid,line width = 1.0pt](0.,2.5mm) -- (5.5mm,2.5mm)}}}

\newcommand{\greendashedline}{\raisebox{0pt}{\tikz{\draw[-,green,dashed,line width = 1.0pt](0.,2.5mm) -- (5.5mm,2.5mm)}}}

\newcommand{\PineGreenline}{\raisebox{0pt}{\tikz{\draw[-,PineGreen,solid,line width = 1.0pt](0.,2.5mm) -- (5.5mm,2.5mm)}}}

\newcommand{\bluecirclee}{\raisebox{0pt}{\tikz{\draw[-,Blue,solid,line width = 1.0pt] (2.75mm,0.8mm) circle (0.75mm);\draw[-,Blue,solid,line width = 1.0pt](0.,0.8mm) -- (5.5mm,0.8mm)}}}

\newcommand{\Turquoiseline}{\raisebox{0pt}{\tikz{\draw[-,Turquoise,solid,line width = 1.0pt](0.,2.5mm) -- (5.5mm,2.5mm); \draw[-,Turquoise,loosely dotted,line width = 1.0pt](5.5mm,2.5mm) -- (11.0mm,2.5mm); \draw[-,Turquoise,solid,line width = 1.0pt](11.0mm,2.5mm) -- (16.5mm,2.5mm)}}}

\newcommand{\reddottedline}{\raisebox{0pt}{\tikz{\draw[-,red,loosely dotted,line width = 1.0pt](0.,2.5mm) -- (5.5mm,2.5mm)}}}

\newcommand{\orangedashedline}{\raisebox{0pt}{\tikz{\draw[-,Peach,solid,line width = 1.0pt](0.,2.5mm) -- (2.25mm,2.5mm);\draw[-,Peach,solid,line width = 1.0pt](5.5mm,2.5mm) -- (7.75mm,2.5mm); \draw[-,Peach,solid,line width = 1.0pt](10.0mm,2.5mm) -- (12.25mm,2.5mm)}}}

\newcommand{\Blackdashedline}{\raisebox{0pt}{\tikz{\draw[-,black,solid,line width = 1.0pt](0.,2.5mm) -- (5.5mm,2.5mm); \draw[-,black,solid,line width = 1.0pt](7.75mm,2.5mm) -- (8.75mm,2.5mm); \draw[-,black,solid,line width = 1.0pt](11.0mm,2.5mm) -- (16.5mm,2.5mm)}}}

\begin{frontmatter}

\title{CFD-driven Symbolic Identification of Algebraic Reynolds-Stress Models}


\author[mymainaddress]{Ismaïl BEN HASSAN SAÏDI\corref{mycorrespondingauthor}}
\cortext[mycorrespondingauthor]{Corresponding author. Present affiliation: Laboratoire DynFluid, Arts et M\'etiers Institute of Technology, 151 Boulevard de l'Hopital, 75013 Paris, France}
\ead{ismail.benhassansaidi@ensam.eu}

\author[mysecondaryaddress]{Martin SCHMELZER}

\author[mythirdaddress]{Paola CINNELLA}
\author[myfourthaddress]{Francesco GRASSO}

\address[mymainaddress]{Institut Aérotechnique, CNAM, 15 Rue Marat, 78210 Saint-Cyr-l'École, France}
\address[mysecondaryaddress]{Faculty of Aerospace Engineering, Delft University of Technology, Kluyverweg 2,
Delft, The Netherlands}
\address[mythirdaddress]{Institut Jean Le Rond D'Alembert, Sorbonne Université, 4 Place Jussieu, 75005 Paris, France}
\address[myfourthaddress]{Laboratoire DynFluid, CNAM, 151 Boulevard de l'Hopital, 75013 Paris, France}

\begin{abstract}
A  CFD-driven deterministic symbolic identification algorithm for learning explicit algebraic Reynolds-stress models (EARSM) from high-fidelity data is developed building on
 the frozen-training SpaRTA algorithm of \cite{schmelzer2020machine}. 
 Corrections for the Reynolds stress tensor and the production of transported turbulent quantities of a baseline linear eddy viscosity model (LEVM) are expressed as functions of tensor polynomials selected from a library of candidate functions. The CFD-driven training consists in solving a blackbox optimization problem in which the fitness of candidate EARSM models is evaluated by running RANS simulations. The procedure enables training models against any target quantity of interest, computable as an output of the CFD model. Unlike the frozen-training approach, the proposed methodology is not restricted to data sets for which full fields of high-fidelity data, including second flow order statistics, are available. However,  the solution of a high-dimensional expensive blackbox function optimization problem is required. Several steps are then undertaken to reduce the associated computational burden. First, a sensitivity analysis is used to identify the most influential terms and to reduce the dimensionality of the search space.  Afterwards, the Constrained Optimization using Response Surface (CORS) algorithm, which approximates the black-box cost function using a response surface constructed from a limited number of CFD solves, is used to find the optimal model parameters.  Model discovery and cross-validation is performed for three configurations of 2D turbulent separated flows in channels of variable section using different sets of training data to show the flexibility of the method. The discovered models are then applied to the prediction of an unseen 2D separated flow with higher Reynolds number and different geometry. The predictions of the discovered models for the new case are shown to be not only more accurate than the baseline LEVM, but also of a multi-purpose EARSM model derived from purely physical arguments. The proposed deterministic symbolic identification approach constitutes a promising candidate for building accurate and robust RANS models customized for a given class of flows at moderate computational cost.
\end{abstract}

\begin{keyword}
\texttt{elsarticle.cls}\sep \LaTeX\sep Elsevier \sep template
\MSC[2010] 00-01\sep  99-00
\end{keyword}

\end{frontmatter}


\section{Introduction}
Computational Fluid Dynamics (CFD) simulations of turbulent flows for industrial applications largely rely on the Reynolds-Averaged Navier-Stokes (RANS) equations supplemented with linear eddy viscosity models (LEVM) for the Reynolds stresses. 
Despite their generally inaccurate predictions of non-equilibrium turbulent flow with strong pressure gradients, streamline curvature, anisotropy and separation \cite{wilcox2006turbulence,hellsten2009explicit} LEVM are usually preferred to more sophisticated models --namely, Reynolds stress models (RSM)-- due to their superior robustness and lower computational cost. RSM involve additional transport equations for the Reynolds stresses \cite{launder1975progress,pope2001turbulent,versteeg2007introduction}, which in principle can be derived exactly from the Navier--Stokes equations.
Unfortunately, this process introduces a number of additional unclosed terms (pressure-strain correlations, diffusion and dissipation rates) that need suitable mathematical modeling, as well as the calibration of the associated closure coefficients. The predictive performance and the numerical robustness of RSM is sensitive to the modeling of these terms. 
Furthermore, RSM involve the solution of seven additional transport equations (for 3D flows), thus increasing computational cost compared to LEVM. 

A trade-off between LEVM  and RSM consists in the development of so-called generalized eddy viscosity models \cite{pope1975more}. Within this class of models, Explicit Algebraic Reynolds Stress Models (EARSM) generalize the linear eddy viscosity concept by assuming that the anisotropic part of the Reynolds stress
not only depends on the mean strain rate tensor $S_{ij}$ but also on the rotation rate tensor $\Omega_{ij}$ through a non linear relationship. Specifically, 
Reynolds stress anisotropy is written as a tensor polynomial of $S_{ij}$ and $\Omega_{ij}$ \cite{pope1975more}, the coefficients of which can be determined from the RSM equations simplified under local equilibrium   \cite{Gatski1993} as functions of tensor invariants.
In this sense, classical EARSM partly inherit from modeling hypothesis used for the development of RSM. 
Furthermore, they also require the calibration of a number of closure coefficients, which is usually done for a set of simple canonical flows.
Due to the uncertainties associated with the choice of both the mathematical structure and the closure coefficients the EARSM, as well as other RANS models, are not universally valid and may perform poorly for flow configurations very dissimilar to the calibration ones. The reader may refer to \cite{XiaoCinnella2019} for a discussion of RANS modeling uncertainties.

Several recent studies have illustrated the potential of data-driven approaches for the development of turbulence models, customized for specific classes of flows (see, e.g., \cite{duraisamy2019turbulence,XiaoCinnella2019}). 
Such approaches use machine learning techniques to infer the functional form of the Reynolds stress tensor from high-fidelity experimental or, most often, numerical data. 
While many of them use black-box machine learning such as Gaussian processes, Deep Neural Networks or Random Forests  \cite{parish2016paradigm, ling2016reynolds, wu2018physics-informed} to reconstruct or correct the Reynolds stress tensor, some others have considered open-box machine learning techniques, relying on the use of a set of candidate solutions or function dictionaries to build physically interpretable turbulence models similar to EARSM.
In \cite{weatheritt2016novel}, this goal is achieved by using
non deterministic symbolic regression based on an evolutionary algorithm (Gene Expression Programming, GEP). 
The GEP is efficient in constructing models with improved accuracy for separated flows. Moreover, learned models involving only a few nonlinear terms are found to exhibit lower training and prediction error and higher numerical robustness \cite{weatheritt2017development,akolekar2019development}.
However, the random nature of the search algorithm discovers a model with a different mathematical form at each run, using the same training data. 
A deterministic symbolic identification algorithm called SpaRTA (Sparse Regression of Turbulent Stress Anisotropy) has been proposed
 in \cite{schmelzer2020machine}. SpaRTA combines functions from a predefined library of candidates without any random recombination. The algorithm constrains the search towards algebraic models of minimal complexity using sparsity-promoting regression techniques \cite{brunton2016discovering,rudy2017data}. This method has also shown its efficiency in improving RANS models for the simulation of separated flows behind steps or in convergent-divergent channels. Recently, SpaRTA has been coupled with LES in a multi-fidelity optimization framework, allowing significant improvements over designs based on a simple LEVM  \cite{zhang2021customized}. 
 A similar deterministic identification approach has been also proposed in \cite{beetham2020formulating}, whereby unclosed tensors in the RANS or Reynolds-stress transport equations are written as linear combinations of a tensor basis, whose coefficients are learned from synthetic or DNS data of simple flows using $l_1$-constrained mean square regression. The results are compared to a constitutive law using a similar decomposition, but with coefficients built on Neural Networks (NN) \cite{ling2016reynolds}.
 Because the deterministic identification methodology seeks to uncover underlying physics, far less data is required to achieve reasonable learned models.
As a consequence, it represents potentially an excellent modeling construct for sparse datasets, such as those available from experiments.
 
A limitation of the above-mentioned symbolic identification algorithms relies in their training procedure. The latter uses a frozen-field (or \emph{a priori} \cite{duraisamy2021perspectives}) mode, whereby  frozen mean-field and Reynolds stress high-fidelity data are used to train the constitutive equation for the Reynolds stresses. Once the training is completed, the resulting model is plugged in a CFD solver and used to predict a new flow. Such a strategy has the advantage of simplicity and extremely low cost (no CFD solves are required for the training procedure). The drawbacks are: 1) the need for full-field, high quality data for first and second-order flow statistics (namely, the Reynolds stresses); the latter are generally available only for simple geometries and low to moderate Reynolds numbers, for which DNS or well-resolved LES are possible; 2) due to the offline training, the learned model may cause numerical stiffness once coupled with a CFD code, requiring the introduction of relaxation coefficients \cite{schmelzer2020machine} or implicit treatment to stabilize the CFD calculations and improve their conditioning \cite{beetham2020formulating}; 3) offline training using frozen high-fidelity fields does not guarantee consistency with the system of RANS equations (see \cite{duraisamy2021perspectives} for a more detailed discussion). 

The above-mentioned limitations have motivated recent work toward the development of so-called model-consistent or CFD-driven training algorithms \cite{holland2019towards,zhao2020rans,strofer2021end-to-end}.
In such an approach, the CFD model is embedded within the training process. This is expected to enable flexible use of incomplete or noisy training data for any flow quantity at hand. Moreover as each candidate model is evaluated through a RANS simulation, models preventing the convergence of the solution can be eliminated during the training process. Hence, learned models are expected to be robust. Finally, the CFD-driven approach ensures coupling of the mean field and turbulent transport equations at any stage of the training process, thus guaranteeing model consistency. However, the price to pay is a much higher computational cost of the training procedure, due to the multiple CFD solves required to evaluate model fitness.

In the present paper, we propose a CFD-driven extension of the SpaRTA algorithm \cite{schmelzer2020machine} that evaluates the fitness of candidate models by running RANS simulations. 
The new method generates data-driven EARSM of the same general form of the frozen-training SpaRTA algorithm. Unlike the latter,  however, it is applicable to configurations for which full fields of first and second-order turbulent statistics are not available (for example, experimental databases), as it can be trained
 with any kind of observable data. Examples include incomplete velocity or pressure fields, wall distributions of pressure or skin friction coefficients, or integral quantities such as aerodynamic coefficients. Such data can be made available through  experiments in addition to simulations,  opening the way to the development of data-driven models for complex, high-Reynolds flows.
 In order to reduce the computational burden associated with model training in a highly dimensional parameter space, a preliminary local sensitivity analysis is used to identify the most influential parameters.
Subsequently, a pre-screening of candidate models is carried out using the Reynolds stress anisotropy barycentric map representation \cite{banerjee2007presentation}, so that parameter ranges leading to violation of the realizability conditions are discarded.
Finally, a response surface methodology \cite{regis2005constrained}  is used to construct an inexpensive surrogate of the costly black-box function from a moderate number of CFD evaluations. An approximate solution to the optimization problem is then computed by running the optimizer on the response surface.
As in the frozen-training SpaRTA approach, the turbulence models obtained at the end of the procedure are data-driven EARSM models involving a subset of the initial library of candidate functions.
Numerical experiments for selected separated flows are used to illustrate the accuracy and the generalization capabilities of the models.

The paper is organized as follows. In Section 2, we briefly recall the formulation of LEVM and EARSM models of turbulence. Datasets used for model training and validation are described in Section 3. In Section 4, we recall the main features of the frozen-training SpaRTA algorithm.
Section 5 is devoted to numerical results and discussion. Conclusions and perspectives for future work are provided in Section 6.

\section{Reynolds-Averaged Navier--Stokes equations and turbulence models}

The incompressible RANS equations read:

\begin{equation}
\frac{\partial \overline{u_i}}{\partial x_i} = 0
\label{eq:RANS_mass_conservation_equation}
\end{equation}   

\begin{equation}
\frac{\partial \overline{u}_i}{\partial t} + \frac{\partial (\overline{u}_i \overline{u}_j)}{\partial x_j}  = - \frac{1}{\rho} \frac{\partial \overline{p}}{\partial x_i} + \frac{\partial }{\partial x_j } 
\left(\nu \frac{\partial \overline{u}_i}{\partial x_j} - \overline{ u_i^\prime  u_j^\prime}\right),
\label{eq:RANS_momentum_equation}
\end{equation}
where $\rho$ is the constant density, $\nu$ the fluid kinematic viscosity, $\overline{u_i}$ and $\overline{p}$ the averaged velocity components and pressure and $u_i^\prime$ the velocity fluctuations. \\

The unclosed term $\tau_{ij}=\overline{ u_i^\prime  u_j^\prime}$ is the kinematic Reynolds stress tensor. It can decomposed into an isotropic part $\frac{2}{3} k \delta_{ij}$ and an anisotropic part $a_{ij} = 2k b_{ij}$ (with $b_{ij}$ a non-dimensional anisotropy tensor), such that:
\begin{equation}
\tau_{ij}=\overline{ u_i^\prime  u_j^\prime}=
2k (b_{ij} + \frac{1}{3} \delta_{ij}),
\label{eq:Reynolds_stress_tensor}
\end{equation}
where $k=tr(\tau_{ij})/2$ is the turbulent kinetic energy.
Therefore, the modeling of the entire range of turbulent scales in the RANS approach reduces to the modeling of $k$ and of the anisotropy tensor.
This is generally done by introducing a constitutive relation for $b_{ij}$, i.e. a functional form relating its components to the mean field.

\subsection{Boussinesq analogy and LEVM models}

Most turbulence models currently used in industrial codes are LEVM. 
Such models rely on the so-called Boussinesq analogy, whereby the anisotropic part of $\tau_{ij}$ is written as a linear function of the mean rate-of-strain $S_{ij}=\displaystyle \frac{1}{2} \left(\frac{\partial \overline{u}_i}{\partial x_j}+ \frac{\partial \overline{u}_j}{\partial x_i}\right)$:
\begin{equation}  
 b_{ij}=- \frac{\nu_t}{k}  S_{ij}
 \label{eq:Boussinesq_Analogy}
\end{equation}	
with $\nu_t$ the eddy viscosity coefficient.\\

The LEVM constitutive relation \ref{eq:Boussinesq_Analogy} is then closed by expressing $\nu_t$ as a function of the mean flow.
 Most often, $\nu_t$ is computed using a transport-equation model such as the $k-\omega$ SST \cite{Menter1994}, used here as the baseline model. 
For completeness, the SST $k-\omega$ equations are recalled in \ref{appendix:k-omega_SST}. Although this model is believed to be one of the most accurate LEVM models on average, it still shows limitations in providing reliable results for flows with separation, streamline curvature or strong pressure gradients \cite{wilcox2006turbulence,hellsten2009explicit}.

\subsection{Augmented Boussinesq analogy and EARSM models}

To overcome the limitations of LEVM models, more sophisticated approaches have been developed, such as the generalized eddy viscosity models \cite{pope1975more}. In particular, EARSM generalize the linear eddy viscosity concept by assuming that the anisotropy of the Reynolds-stress depends not only
on the strain rate tensor $S_{ij}$ but also on the rotation rate tensor $\Omega_{ij}=\displaystyle\frac{1}{2} \left(\frac{\partial \overline{u}_i}{\partial x_j}- \frac{\partial \overline{u}_j}{\partial x_i}\right)$ through a non linear relationship.\\

Such models can be constructed from a baseline LEVM using the so-called augmented Boussinesq analogy: 
\begin{equation}
b_{ij}= b_{ij}^{bl} + b^\Delta_{ij},
\label{eq:augmented_boussinesq}
\end{equation}
where $b^{bl}_{ij}=-\displaystyle\frac{\nu_t}{k} S_{ij}$ and $b^\Delta_{ij}$ is the extra anisotropy tensor, used to correct the baseline Boussinesq model $b_{ij}^{bl}$.\\
Following this approach, an EARSM can be constructed by specifying a non linear relationship between the extra anisotropy tensor $b^\Delta_{ij}$ and the strain rate tensor and the rotation rate tensor.
Following \cite{pope1975more}, the most general form of the anisotropic Reynolds-stress correction is derived via the Cayley-Hamilton theorem:  
\begin{equation}
b^\Delta_{ij} ( \hat{S}_{ij} , \hat{\Omega}_{ij})=\sum_{\lambda=1}^{10} G_\lambda (I_1,...,I_5) T^{(\lambda)}_{ij},
\label{eq:combinaison_lineaire}
\end{equation}
using a minimal integrity basis of ten tensors $T^{(\lambda)}_{ij}$ and five invariants $I_m$, written as functions of the non-dimensional mean strain rate $\hat{S}_{ij}=\displaystyle\frac{1}{2\omega}\left(\frac{\partial \overline{u}_i}{\partial x_j}+ \frac{\partial \overline{u}_j}{\partial x_i}\right)$ and rotation rate $\hat{\Omega}_{ij}=\displaystyle\frac{1}{2\omega}\left(\frac{\partial \overline{u}_i}{\partial x_j}- \frac{\partial \overline{u}_j}{\partial x_i}\right)$. 
In the following, we focus on two-dimensional flow cases, for which the first three base tensors form a linearly independent basis and only the first two invariants are nonzero \cite{pope1975more}, and the set of base tensors and invariants reads

\begin{equation}
\begin{array}{l l}
T_{ij}^{(1)} &= \hat{S}_{ij},\; T_{ij}^{(2)} = \hat{S}_{ik}\hat{\Omega}_{kj} - \hat{\Omega}_{ik}\hat{S}_{kj}, \nonumber \\
T_{ij}^{(3)} &= \hat{S}_{ik}\hat{S}_{kj} - \frac{1}{3} \delta_{ij} \hat{S}_{mn}\hat{S}_{nm}\\
I_1 &= \hat{S}_{mn}\hat{S}_{nm},\; I_2 = \hat{\Omega}_{mn}\hat{\Omega}_{nm}.
\end{array}
\label{eq:tensorial_basis}
\end{equation} 

Deriving EARSM models then consists in finding a set of scalar functions $G_\lambda$, which are classically obtained from physical considerations \cite{pope1975more,Gatski1993}. In the following, we present a deterministic symbolic identification method to build EARSM models from high-fidelity data sets, cast in the form of equations (\ref{eq:Reynolds_stress_tensor}), (\ref{eq:augmented_boussinesq}) and (\ref{eq:combinaison_lineaire}).  

In the numerical experiments of section 5, we also consider a general purpose EARSM, recently proposed in \cite{menter2012explicit} and referred-to as BSL-EARSM, for comparison with customized data-driven models. Its formulation is briefly recalled in \ref{appendix:BSL}.

\subsection{CFD solver}
The CFD-SpaRTA methodology requires the implementation of the general formulation of the augmented EARSM in a CFD solver. In the present work, the open-source finite-volume code OpenFOAM \cite{weller1998tensorial} has been used. The RANS equation system (equations \ref{eq:RANS_mass_conservation_equation} and \ref{eq:RANS_momentum_equation}) are solved using the SIMPLE algorithm \cite{caretto1973two}; the convective terms are discretized using linear upwinding and viscous terms with 2nd order central differencing. 
The solution is advanced to the steady state using a Gauss-Seidel smoother.

\section{Flow configurations and data sets}
\label{sec:datasets}
The aim of the present work being the development of data-driven corrections to turbulence models, all numerical experiments are carried out for a family of flow configurations that challenge standard RANS models, and specifically LEVM.
In particular, we have considered the same cases reported in \cite{schmelzer2020machine}, representing flows in 2D channels with variable cross sections.
High fidelity data from DNS or well-resolved LES are available for model training and validation.   
To ensure that discretization errors remain low with respect to turbulence modeling errors, the RANS simulation use sufficiently fine meshes. 

More details of the flow configurations and data sets used in the present numerical experiments presented are given below.

\subsection{Periodic hill flow (PH$_{10595}$)}	
The so-called periodic hill flow consists of a flow through a channel constrained by periodic restrictions (hills) of height $h$. For a channel segment comprised between two adjacent hills, the flow separates on the lee-side of the first hill and reattaches between the hills. The test case has been widely investigated in the literature, both experimentally and numerically.
The high-fidelity LES data used in the present work are from \cite{breuer2009flow} for $Re = 10595$ ($\textrm{PH$_{10595}$}$), where Re is a Reynolds number based on the bulk velocity in the restricted section and the hill height. Our RANS simulations use a computational grid consisting of 120 x 130 cells. Cyclic boundary conditions are used at the inlet and outlet and a forcing term is applied to maintain a constant flow rate through the channel.

\subsection{Converging-diverging channel at Re=12600 (CD$_{12600}$)}
This configuration corresponds to a 2D channel of half-height $H$ with an asymmetric bump of height $h\approx 2/3 H$ located on the bottom wall. 
The Reynolds number (based on the channel half-height and inlet conditions) is $Re=12600$. A small separation occurs on the lee-side of the bump. For this test case,  high-fidelity DNS data from \cite{marquillie2011instability} are available. 
The RANS simulations are based on a mesh of 140 x 100 cells. 
A velocity profile obtained from a companion channel-flow simulation is imposed at the inlet of the computational domain.

\subsection{Curved backward-facing step (CBFS$_{13700}$)}
The case consists in a 2D flow over a gently-curved backward-facing step of height $h$, producing a separation bubble. 
The upstream channel height is 8.52$h$ and the Reynolds number, based on the inlet velocity and step height, is 13700.
High-fidelity LES data from \cite{bentaleb2012large} are used for training. For the RANS simulations, the mesh consists of 140 x 150 cells. Slip conditions are used at the upper boundary, and a velocity profile obtained from a fully-developed boundary layer simulation is set at the domain inlet.   

\section{Symbolic identification methodology}
In this section, we introduce the general form of candidate models, constructed from a library of functions. Then we recall the frozen-training SpaRTA algorithm of \cite{schmelzer2020machine}. 
Finally, we describe in detail the steps of the new CFD-driven symbolic identification algorithm.

\subsection{Symbolic identification of EARSM} \label{symbolic}
Following the approach proposed in \cite{schmelzer2020machine}, the scalar functions $G_\lambda$ are approximated as linear combinations of candidate functions of 
$I_1$ and $I_2$, consisting of monomials up to order 6:
\begin{equation}
\begin{array}{l}
\mathbf{B}=[c, I_1,I_2,I_1^2, I_1I_2,I_2^2,I_1^3, I_1^2I_2, I_1 I_2^2,I_2^3, I_1^4, I_1^3I_2, I_1^2 I_2^2,I_1 I_2^3,\\ I_2^4, I_1^5, I_1^4 I_2, I_1^3 I_2^2, I_1^2 I_2^3, I_1 I_2^4, I_2^5, I_1^6, I_1^5 I_2, I_1^4 I_2^2, I_1^3 I_2^3, I_1^2 I_2^4, I_1 I_2^5, I_2^6 ]^T.
\end{array}
\label{eq:schmelzer_library}
\end{equation}  

\noindent Given this library, the full expression of eq. (\ref{eq:combinaison_lineaire}) becomes:

\begin{equation}
\begin{array}{l l}
b^\Delta_{ij} ( S_{ij} , \Omega_{ij}) &= \theta^{b^\Delta_{ij}}_0 T_{ij}^{(1)} + \theta^{b^\Delta_{ij}}_1 I_1 T_{ij}^{(1)} + \theta^{b^\Delta_{ij}}_2 I_2 T_{ij}^{(1)} + \dots \\
&+ \theta^{b^\Delta_{ij}}_{26} I_1 I_2^5 T_{ij}^{(1)} + \theta^{b^\Delta_{ij}}_{27} I_2^6 T_{ij}^{(1)}  \\
&+ \theta^{b^\Delta_{ij}}_{28} T_{ij}^{(2)} + \theta^{b^\Delta_{ij}}_{29} I_1 T_{ij}^{(2)} + \theta^{b^\Delta_{ij}}_{30} I_2 T_{ij}^{(2)} + \dots \\
&+ \theta^{b^\Delta_{ij}}_{54} I_1 I_2^5 T_{ij}^{(2)} + \theta^{b^\Delta_{ij}}_{55} I_2^6 T_{ij}^{(2)}  \\
&+ \theta^{b^\Delta_{ij}}_{56} T_{ij}^{(3)} + \theta^{b^\Delta_{ij}}_{57} I_1 T_{ij}^{(3)} + \theta^{b^\Delta_{ij}}_{58} I_2 T_{ij}^{(3)} + \dots \\
&+ \theta^{b^\Delta_{ij}}_{82} I_1 I_2^5 T_{ij}^{(3)} + \theta^{b^\Delta_{ij}}_{83} I_2^6 T_{ij}^{(3)}, 
\end{array}
\end{equation} 
\noindent where the vector of coefficients: 
\begin{equation}
\Theta^{b^\Delta_{ij}}  = \Big[ \theta^{b^\Delta_{ij}}_0, \theta^{b^\Delta_{ij}}_1, \theta^{b^\Delta_{ij}}_2, \dots , \theta^{b^\Delta_{ij}}_{83} \Big]^T\quad\in{\mathbb R}^{84}
\end{equation} 
\noindent has to be determined.

A major improvement proposed in \cite{schmelzer2020machine} consists in the introduction of a corrective term in the turbulent transport equations, in addition to the use of the generalized eddy viscosity formulation. Specifically, a residual term $R$ is added to the model transport equations and the augmented k-$\omega$-SST model becomes:  

\begin{equation}
\frac{\partial k}{\partial t} + \overline{u}_j \frac{\partial k}{\partial x_j} = P_k + R -\beta ^ * k \omega + \frac{\partial }{\partial x_j} [(\nu + \sigma_k \nu_t) \frac{\partial k}{\partial x_j}]
\label{eq:schmelzer_k_transport}
\end{equation}

\begin{equation}
\begin{array}{l}
\displaystyle\frac{\partial \omega}{\partial t} + \overline{u}_j \frac{\partial \omega}{\partial x_j} =  \frac{\gamma}{\nu_t} (P_k + R) - \beta \omega^2 + 
\displaystyle \frac{\partial }{\partial x_j} [(\nu + \sigma_\omega \nu_t) \frac{\partial \omega}{\partial x_j}] + C D_{k \omega},
\end{array}
\label{eq:schmelzer_omega_transport}
\end{equation} 
with $P_k=\textrm{min}(-2k(b^{bl}_{ij}+b^\Delta_{ij}) \partial_j U_i,  10 \beta^*\omega k)$.

The residual $R$ can be positive or negative and it represents a correction of the model production term.
Introducing a suitable non-dimensional correction tensor $b_{ij}^R$, the model for $R$ writes:
\begin{equation}
R = 2kb^R_{ij} \frac{\partial \overline{u}_j}{\partial x_i},   
\label{eq:R_modelling_assumption}   
\end{equation} 
where $b^R_{ij}$ is also decomposed according to the Cayley-Hamilton theorem:

\begin{equation}
\begin{array}{l l}
b^R_{ij} ( S_{ij} , \Omega_{ij}) &= \theta^R_0 T_{ij}^{(1)} + \theta^R_1 I_1 T_{ij}^{(1)} + \theta^R_2 I_2 T_{ij}^{(1)} + \dots \\
&+ \theta^R_{26} I_1 I_2^5 T_{ij}^{(1)} + \theta^R_{27} I_2^6 T_{ij}^{(1)}  \\
&+ \theta^R_{28} T_{ij}^{(2)} + \theta^R_{29} I_1 T_{ij}^{(2)} + \theta^R_{30} I_2 T_{ij}^{(2)} + \dots \\
&+ \theta^R_{54} I_1 I_2^5 T_{ij}^{(2)} + \theta^R_{55} I_2^6 T_{ij}^{(2)} \\
&+ \theta^R_{56} T_{ij}^{(3)} + \theta^R_{57} I_1 T_{ij}^{(3)} + \theta^R_{58} I_2 T_{ij}^{(3)} + \dots \\
&+ \theta^R_{82} I_1 I_2^5 T_{ij}^{(3)} + \theta^R_{83} I_2^6 T_{ij}^{(3)}, 
\end{array}
\end{equation}
\noindent The vector of coefficients:

\begin{equation}
\Theta^{R}  = \Big[ \theta^{R}_0, \theta^{R}_1, \theta^{R}_2, \dots , \theta^{R}_{83} \Big]^T\quad\in{\mathbb R}^{84}
\end{equation} 

\noindent needs to be determined alongside with $\Theta^{b_{ij}^\Delta}$.

Once the general library of candidate models has been chosen, the difference between the frozen-training and the CFD-driven approaches relies in the method used to determine $\Theta^{b^\Delta_{ij}}$ and $\Theta^{R}$.

\subsection{Frozen-training SpaRTA algorithm}
\label{subsec:frozen-training}
In this approach, high fidelity data for the velocity field, $\overline{u}_i^*$, turbulent kinetic energy $k^*$ and Reynolds-stresses $\tau_{ij}^*$ are used to train the model. Such data are pre-processed prior to use for model learning. 

The high-fidelity anisotropy tensor
$b_{ij}^*$ is obtained from equation (\ref{eq:Reynolds_stress_tensor}). 
The next step of the method is the determination of the high fidelity values for $b^\Delta_{ij}$ and $R$. Knowing $b_{ij}^*$, $b^{\Delta,*}_{ij}=b^*_{ij}+\frac{\nu_t^*}{k^*} S^*_{ij}$ is directly derived if a high-fidelity ansatz of the turbulent viscosity $\nu^*_t$ is provided. 
The latter can be computed using the relation (\ref{eq:k-omega-SST_nut}), setting $k=k^*$. However, a high fidelity value of $\omega$ is also needed. 
For this purpose, a so-called $k$-corrective-frozen-RANS algorithm is used, whereby a high-fidelity estimate of the residual $R^*$ is computed alongside $\omega^*$. This process consists in injecting $\overline{u_i}^*$, $k^*$ and $b_{ij}^*$ in equations (\ref{eq:schmelzer_k_transport}) and (\ref{eq:schmelzer_omega_transport}) and iteratively computing $\omega^*$ by solving equation (\ref{eq:schmelzer_omega_transport}). At each iteration, $R^*$ is computed as the residual of equation (\ref{eq:schmelzer_k_transport}) and fed back into equation (\ref{eq:schmelzer_omega_transport}). 
On completion of the preceding manipulations, high fidelity fields for $b^{\Delta,*}_{ij}$ and $R^*$ are available. 

High-fidelity counterparts of the terms in the library $\mathbf{B}^*$ are computed from the high-fidelity velocity field by using relations (\ref{eq:tensorial_basis}); 
similarly, high-fidelity estimates of the basis tensors $T^{(\lambda,*)}_{ij}$ are computed. $\mathbf{B}^*$ and $T^{(\lambda,*)}_{ij}$ are then used to compute the $b^\Delta_{ij}$ and $R$ fields corresponding to candidate models defined by specifying the sets of coefficients $\Theta^{b^\Delta_{ij}}$ and $\Theta^{R}$. 
For each field $\Delta$ ($\Delta$ indicating either $b^\Delta_{ij}$ or $R$) the following optimization problem can be constructed:
\begin{equation}
\Theta =   \underset{\hat{\Theta}}{\operatorname{argmin}}      \| \Delta^{*} - \Delta( \hat{\Theta}) \|^2_2 + \textrm{regularization terms},
\label{eq:CFD-free_optimization_problem}
\end{equation}
\noindent where $\|\bullet\|_2$ stands for the $l_2$ norm of the error, $\hat{\Theta}$ is the vector $\Theta^{b^\Delta_{ij}}$ (respectively $\Theta^{R}$) if $\Delta$ represents $b^\Delta_{ij}$ (respectively $R$). $\Delta^{*}$ is the high fidelity value of the field and $\Delta(\hat{\Theta})$ is the field reconstructed using the model corresponding to $\hat{\Theta}$. 

The choice of the regularization terms is crucial to efficiently select a small subset of relevant terms among those constituting the highly dimensional symbolic library. The goal of the whole methodology being the construction of augmented RANS models optimized for a specific class of flows, the learnt model should be general enough to be effective for the simulation of flows belonging to the same class (e.g. flows with separations). Without the regularization terms, the problem formulated by equation (\ref{eq:CFD-free_optimization_problem}) reduces to a standard mean-squares optimization problem whose resolution leads to dense coefficient vectors $\Theta^{b^\Delta_{ij}}$ and $\Theta^{R}$, possibly overfitting 
the data.
In \cite{schmelzer2020machine}, this effect is mitigated by using an elastic net regularization, known to be sparsity promoting \cite{brunton2019data,mcconaghy2011ffx} :
\begin{equation}
	\textrm{regularization terms}=\lambda \rho \| \hat{\Theta} \|_1 + 0.5 \lambda (1- \rho) \| \hat{\Theta} \|_2^2,
	\label{eq:elastic_net}
\end{equation}
\noindent where $\rho\in [0,1]$ is the blending parameter between the $l_1-$ and $l_2-$ norm regularization and $\lambda$ is the regularization weight. The blending parameter $\rho$ relaxes the $l_1-$ regularization, that tends to select only one function in sets of correlated functions and therefore degrade the predictive performance of the inferred models. This relaxation allows to find sparse models with good predictive performance. The regularization weight $\lambda$ also controls the sparsity of the inferred models (independently of $\rho$). As the optimal value of $(\rho,\lambda)$ is not known a priori, the optimization problem given in equation (\ref{eq:CFD-free_optimization_problem}) must be solved for a grid of pairs $(\rho_{1 \leq i \leq N},\lambda_{1 \leq j \leq M})$. In Schmelzer et al. \cite{schmelzer2020machine}, a set of optimizations is carried out a grid of $9\times 100$, leading to the discovery of 900 concurrent models, some of which have the same abstract form (i.e. involve the same functional terms). Once the abstract model forms have been identified, an inference step based on ridge regression is used to calibrate the model coefficients. The computational cost of model discovery and calibration is of the order of minutes, since no additional CFD solve is required at this stage. Finally, best-performing models are selected from the wide set of learned models through cross validation \cite{bishop2006pattern}. In order to not overcharge the
role of the training data from k-corrective-frozen-RANS, the validation task is performed \emph{a posteriori} by implementing candidate models in a CFD solver. This allows validation against quantities other than those used in the training process, namely velocity data. Specifically, data for cases PH$_{10595}$, CD$_{12600}$ and CBFS$_{13700}$ are used, leading to approximately 200 CFD calculations.

The CFD solver used in \cite{schmelzer2020machine} is OpenFOAM \cite{weller1998tensorial}, in which the general formulation of the augmented EARSM has been implemented as a new class of models. The coefficients are passed by the optimizer to the OpenFoam input file \url{RASproperties} using a python interface.
The code of the frozen-training SpaRTA has been made public and can be found at the following github repository : \href{https://github.com/shmlzr/general_earsm.git}{https://github.com/shmlzr/general\_earsm.git}.

\subsection{CFD-driven symbolic identification}
\label{par:CFD-driven symbolic identification}
\begin{figure}[H]
	\centering
	\centerline{\includegraphics[width=1\linewidth]{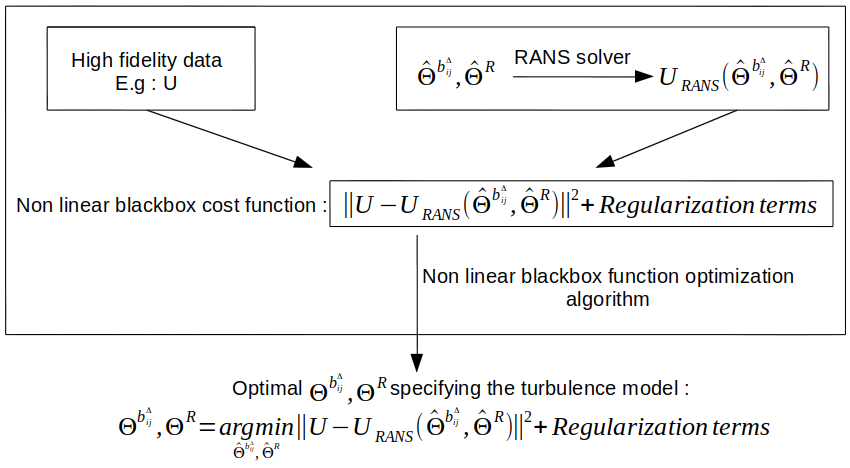}}
	\caption{Technical flow diagram of the CFD-driven SpaRTA method. $U$ represents any quantity of interest considered as the target for model training.}
	\label{fig:CFD-driven-sparta_diagram}
\end{figure}

The technical flow diagram of the CFD-driven algorithm that we have developed in the present work to determine $\Theta^{b^\Delta_{ij}}$ and $\Theta^{R}$
 is sketched in figure \ref{fig:CFD-driven-sparta_diagram}.

The first step of the method consists in gathering high fidelity data for a flow or a class of flows for which we want to improve the accuracy of the baseline model. It should be emphasized that, while the frozen-training approach requires data for second-order quantities like $k$ and $\tau_{ij}$, observations available for any quantity of interest
(velocity or pressure measurements, global aerodynamic coefficients, etc.)  can be used as a target for training.
In the flowchart of figure \ref{fig:CFD-driven-sparta_diagram}, we assumed that high fidelity data are available for a quantity of interest $U$: this could be, for instance, the streamwise velocity component $\overline{u}_1$ but the same procedure applies for any other data set. Typically, $U$ is a vector of high-fidelity samples extracted at selected locations in the flow field and concatenated into a single vector.\\ 

In the second step, the general form of the augmented turbulence model is implemented in the CFD-solver, as described in Section \ref{subsec:frozen-training}.
The latter is used to evaluate the quantity $U$ using a candidate augmented turbulence model specified by any instance of the vectors $\hat{\Theta}^{b^\Delta_{ij}}$ and $\hat{\Theta}^{R}$. The output RANS data for $U$ are indicated as $U_{RANS}(\hat{\Theta}^{b^\Delta_{ij}}, \hat{\Theta}^{R})$. Then, the fitness of candidate models is evaluated in terms of the mean squared error $\| U^* - U_{RANS}(\hat{\Theta}^{b^\Delta_{ij}}, \hat{\Theta}^{R}) \|^2_2$ between the high fidelity data $U^*$ and the RANS model prediction $U_{RANS}(\hat{\Theta}^{b^\Delta_{ij}}, \hat{\Theta}^{R})$.
The model can then been constructed by solving the following optimization problem:

\begin{equation}
({\Theta}^{b^\Delta_{ij}}, {\Theta}^{R}) =   \underset{(\hat{\Theta}^{b^\Delta_{ij}}, \hat{\Theta}^{R})}{\operatorname{argmin}}      \| U - U_{RANS}(\hat{\Theta}^{b^\Delta_{ij}}, \hat{\Theta}^{R}) \|^2_2 + \textrm{regularization terms},
\label{eq:CFD-driven_optimization_problem}
\end{equation}

\noindent where regularization terms are added to the least squares formulation to prevent overfitting of the high fidelity data. The choice of the regularization is discussed in the following. 

The preceding cost function is a non-linear black-box function involving a RANS simulation for any choice of $\hat{\Theta}^{b^\Delta_{ij}}, \hat{\Theta}^{R}$, and is therefore costly in terms of CPU time.  
In the following, problem \ref{eq:CFD-driven_optimization_problem} is optimized using the Constrained Optimization using Response Surface (CORS) algorithm \cite{regis2005constrained}, which uses the response surface methodology to approximate the expensive function using a limited number of function evaluations. 
The optimization is subsequently performed on this response surface which is further enriched at each iteration with a new function evaluation at the provisional optimum location (see \ref{appendix:CORS} for a short description of the algorithm). 
Here we use a version of the CORS algorithm adapted from the package available in the \textit{blackbox} python module (see Ref. \cite{knysh2016blackbox}). 

A first modification is introduced to account for diverging or poorly converging CFD simulations due to non realizable/non robust candidate models. If convergence is not achieved according to the prescribed criteria, the candidate model is eliminated and the coefficients are resampled by slightly decreasing their magnitude. The procedure is repeated until convergence is obtained.
Then, the initial response surface is built using $n$ function evaluations, while additional $m$ evaluations are used to converge the solution. The parameters $m$ and $n$ are set independently and typically
$n<m$ and $n+m>2D$ to ensure a good exploration of the parameter space and a satisfactory convergence to the optimum, with $D$ the dimensionality of the search space.
Note that in the original CORS algorithm $n=m=N/2$, with $N>2D$.

Given the difficulty of constructing accurate response surfaces in highly dimensional parameter spaces (here, $D=168$),
a preliminary local sensitivity analysis is conducted to identify the most influential parameters and reduce the search space. 
Afterwards, search ranges for model parameters are assigned based on Reynolds-stress realizability considerations. 
Finally, the regularization terms in the fitness function definition (\ref{eq:CFD-driven_optimization_problem}) are specified.
These three steps are described into more details in the following.

\subsubsection{Local sensitivity analysis}

In the CORS algorithm here implemented, the total number of evaluations $N=n+m$ must satisfy the constraint $N>2D$. As the general formulation of the augmented model involves 168 coefficients, we have $N>336$. In addition to the large numerical cost of the training procedure, for highly dimensional spaces, the response surface may not provide an accurate enough approximation of the true cost function.

To reduce the size of the parameter space, we have carried out a local sensitivity analysis around the baseline model formulation (corresponding to $\theta^{b^{\Delta}}=0$ and $\theta^{R}=0$) to determine the most influential coefficients. For that purpose, each coefficient is varied by a small perturbation ($O(10^{-3})$), and the corresponding CFD solution is determined.
The derivatives of the cost function are then evaluated by one-sided finite differences. 
More precisely, we investigate the sensitivity of the mean squared error on prescribed reference data to parameters $\Theta^{b^\Delta_{ij}}$ and $\Theta^{R}$ of the general EARSM.

The results of the sensitivity study are reported in the following for the three flow configurations $\textrm{PH$_{10595}$}$, CD$_{12600}$ and CBFS$_{13700}$ of Section \ref{sec:datasets}. For the three flows, the cost function is based on high fidelity velocity and Reynolds stress profiles. The profiles are selected at the following streamwise locations: for  PH$_{10595}$ at $x/h=(0.05, 0.5,  1.0, 2.0, 3.0, 4.0, 5.0, 6.0, 7.0, 8.0 )$;
for CD$_{12600}$ at $x/H=(5.5,6.5,7.5,8.5,9.5,10.5,11.5,12)$; and for CBFS$_{13700}$ $x/h=(0.1,1.5,3.0,4.5,6.0,7.5)$.

\begin{figure}[H]	
	\centering
	\begin{subfigure}[b]{1.\textwidth}
		\centering
		\includegraphics[width=0.85\textwidth]{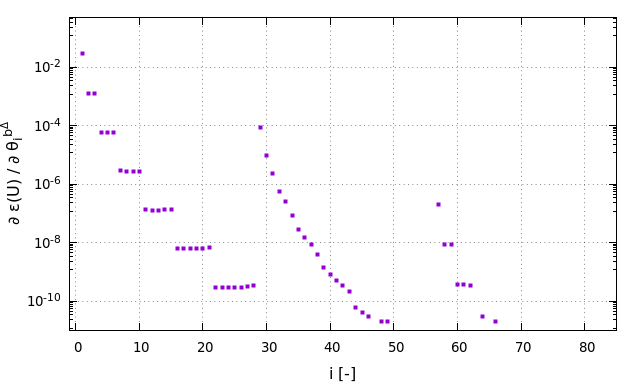}
		\caption{PH$_{10595}$}
		\label{fig:local_SA_PH}
	\end{subfigure}
	\vfill
	\begin{subfigure}[b]{1.\textwidth}
		\centering
		\includegraphics[width=0.85\textwidth]{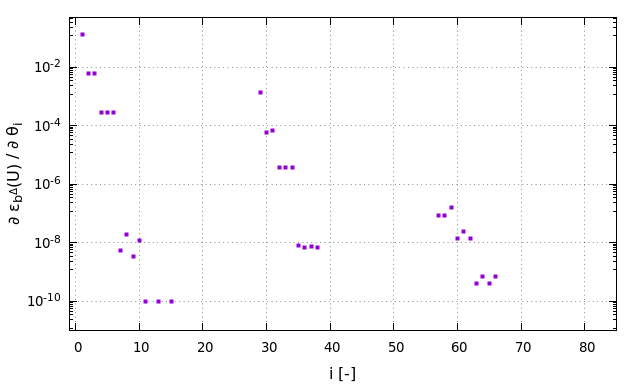}
		\caption{CD$_{12600}$}
		\label{fig:local_SA_CD}
	\end{subfigure}
	\vfill
	\begin{subfigure}[b]{1.\textwidth}
		\centering
		\includegraphics[width=0.85\textwidth]{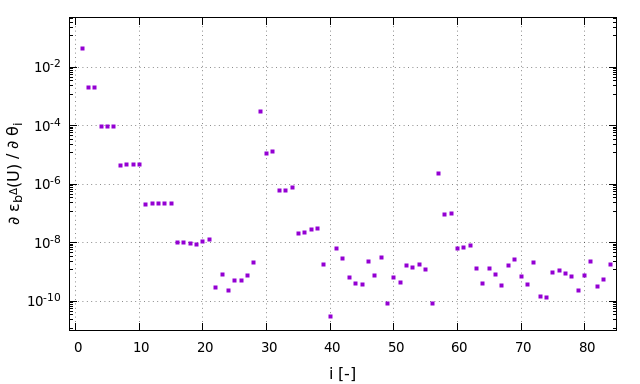}
		\caption{CBFS$_{13700}$}
		\label{fig:local_SA_CBFS}
	\end{subfigure}
	\caption{Sensitivity derivatives of the cost function (based on the velocity components) to the parameters $\theta^{b^{\Delta}}$ for the three test cases.}
	\label{fig:local_SA}
\end{figure}

The sensitivity derivatives of the cost function with respect to the parameters of the $b^{\Delta}$ model are reported in figure \ref{fig:local_SA}, where the abscissas correspond to parameter subscripts.
Parameters with subscripts $0\div 27$, $28\div 55$ and $56\div 83$  appear in the function coefficients of $T^{(1)}_{ij}$, $T^{(2)}_{ij}$ and $T^{(3)}_{ij}$, respectively. Several comments are in order. First, the parameters associated with $T^{(1)}_{ij}$ (corresponding to a correction of the linear eddy viscosity coefficient) are overall more influential than the parameters involved in the quadratic tensorial terms $T^{(2)}_{ij}$ and $T^{(3)}_{ij}$, which govern higher-order corrections of the constitutive law.
Second, the parameter sensitivity decays rapidly with the degree of the monomial functions, and terms up to the second degree are at least one order of magnitude larger than the coefficients of the higher order terms. 
This has led us to retain only monomials up to the second order in the library of candidate functions (equation (\ref{eq:schmelzer_library})). 
Similar results are found for the parameters of $b^{R}$. 
On the basis of this analysis, the dimensionality of the search space is reduced to 36 instead of the initial value 168.

	\subsubsection{Selection of parameter ranges}

	The initial range of each model coefficient must be specified prior to the CORS optimization. For each $\Theta_{1 \leq i \leq 36}$ of $\Theta^{b^\Delta_{ij}}$ and $\Theta^{R}$, the search space is set as:
	\begin{equation}
	a_i\leq \Theta_i \leq b_i
	\label{eq:search_space}
	\end{equation}	
	where $a_i \in \mathbb{R}$ and $b_i \in \mathbb{R}$ are respectively the lower and upper bound of the range of variation of candidate model coefficients. In our setup of the CFD-driven method, the coefficients ranges are set with the aim to ensure the realizability constrain \cite{speziale1994realizability}.
		
    As stressed in Section \ref{par:CFD-driven symbolic identification} and in \ref{appendix:CORS} candidate models are evaluated by running RANS simulations at each iteration of the optimization loop. When a candidate model is found to be non robust, preventing convergence, it is eliminated and resampled until a robust model is obtained. This procedure introduces an additional computational cost that depends on the number of non robust models to be evaluated. This number increases with the dimension of the search space $D$. Hence, the \emph{a priori} specification of parameter ranges results from the following trade-off. On one hand, the parameter range must be large enough to obtain substantial modification of the baseline model. On the other, it must be as small as possible to reduce the number of non robust candidate models and avoid the overcost associated with testing / elimination and resampling.
    
	For that purpose, a realizability study is performed using the Reynolds stress anisotropy barycentric map representation \cite{banerjee2007presentation,emory2014visualizing,edeling2018data}. 
	
  \paragraph{A priori realizability study}
	The modeled anisotropy tensor is initially computed from high fidelity data with each parameter $\Theta^{b^\Delta_{ij}}$
	varying in the range $[-7,7]$.
The tensor eigenvalues are noted $\lambda_1$, $\lambda_2$ and $\lambda_3$ (with $\lambda_1 \geq \lambda_2 \geq \lambda_3$). Following \cite{banerjee2007presentation}, we construct an equilateral triangle defined by the three corners $\mathbf{x}_{1C}$, $\mathbf{x}_{2C}$ and $\mathbf{x}_{3C}$, where $\mathbf{x}=(x,y)$ is a position vector. The three corners the triangle represent the three limiting physical states of turbulence. 
We then construct the following linear mapping between the anisotropy eigenvalues and the coordinates $x$ :
\begin{equation}
\mathbf{x}=\mathbf{x}_{1C}(\lambda_1-\lambda_2)+\mathbf{x}_{2C}(2\lambda_2-2\lambda_3)	+\mathbf{x}_{3C}(3\lambda_3+1).
\label{eq:}
\end{equation}
As any realizable state of turbulence is a convex combination of the three limiting states of turbulence, $\mathbf{x}$ must lie within the triangle if $\lambda_1$, $\lambda_2$ and $\lambda_3$ correspond to a realizable anisotropy tensor. For a given set of coefficients $\Theta^{b^\Delta_{ij}}$, the eigenvalues of $b_{ij}$ can be computed for each mesh point and the corresponding $\mathbf{x}$ can be placed in the barycentric map in order to verify if the resulting anisotropy tensor field corresponds to a physically realizable flow.

	A total of 50000 vectors $\Theta^{b^\Delta_{ij}}$ are generated using uniform Latin Hypercube sampling. 
	For each sample, the corresponding anisotropy tensor eigenvalues are plotted in the barycentric map coordinate system for each mesh point. Sets of parameters leading to a violation of the realizability condition are then discarded. For the remaining sets of parameters, we compute the average value, the standard deviation and the minimum and maximum values taken by each component of $\Theta^{b^\Delta_{ij}}_l$. Realizable values of the coefficients are equal to 0 in the mean, with a variance of approximately 4. 
	Therefore, ranges $[-2,2]$ are finally adopted for the parameters $\Theta^{b^\Delta_{ij}}$ and the same search range is assumed to be appropriate for the  $\Theta^{R}_{ij}$.

\begin{figure}[h]
	\begin{minipage}[c]{0.5\linewidth}
		(a)\includegraphics[width=1.0\linewidth]{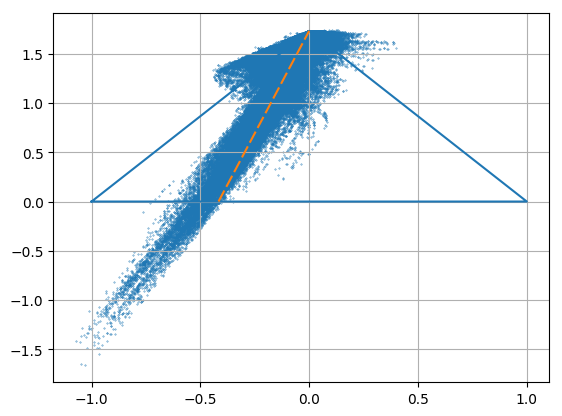}
	\end{minipage} \hfill
	\begin{minipage}[c]{0.5\linewidth}
		(b)\includegraphics[width=1.0\linewidth]{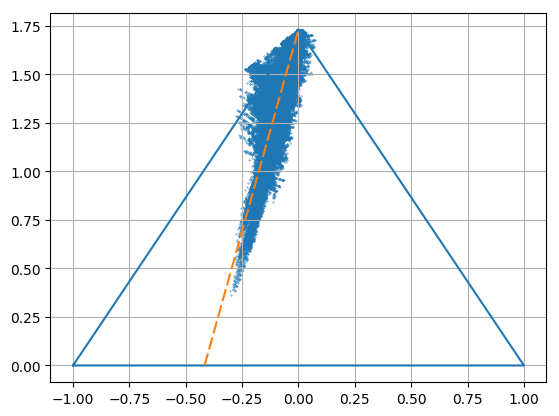}
	\end{minipage}
\caption{Modeled anisotropy tensor at each mesh cell placed in the barycentric map for 100 random vectors $\Theta^{b^\Delta_{ij}}$ sampled using uniform Latin Hypercube sampling. (a) sampling in range $[-7,7]^{18}$. (b) Sampling in range $[-2,2]^{18}$}.
\label{fig:Barycentric_maps}
\end{figure}

In figure \ref{fig:Barycentric_maps}a, results for the PH$_10595$ case are reported. For the sake of clarity, the modeled anisotropy tensor at each mesh cell has been placed  in the barycentric map using only 100 random vectors $\Theta^{b^\Delta_{ij}}$ sampled using uniform Latin Hypercube sampling in range $[-7,7]^{18}$. The figure shows that the anisotropy tensor is not realizable for a large fraction of mesh points. In figure \ref{fig:Barycentric_maps}b, the sampling has been performed in range $[-2,2]^{18}$. The fraction of mesh points possibly corresponding to non realizable anisotropy tensors is drastically reduced. Therefore, an initial parameter range $[-2,2]$ ensures that most candidate models lead to realizable anisotropy tensor fields in the entire computational domain. 
	Generally, this choice yields more robust models and limits the risk non-converging CFD simulations. The remaining non-robust models are excluded from the CORS optimization as mentioned in Section \ref{par:CFD-driven symbolic identification} (see also \ref{appendix:CORS}).

\begin{figure}[H]
	\centering
	 \includegraphics[width=1.0\linewidth]{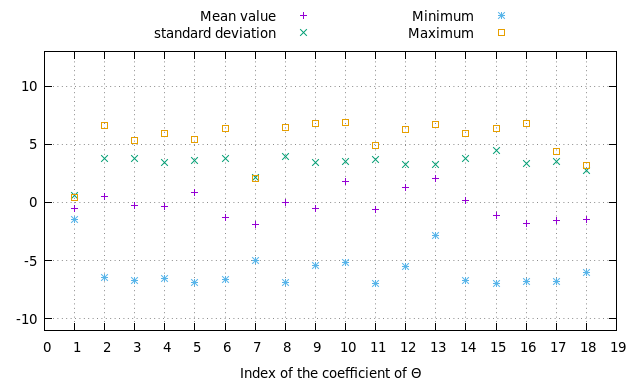} 
	\caption{Mean value, variance, minimum and maximum value of each coefficient $\Theta^{b^\Delta_{ij}}_{{1 \leq l \leq 18}}$ among vector samples corresponding to realizable models.} 
	\label{fig:realizability}
\end{figure}

	\paragraph{CFD-based realizability study} As an alternative to using high-fidelity full field data, a CFD-based realizability study can be performed prior to the optimization step.
   Here, a set of 300 random vectors $\Theta^{b^\Delta_{ij}}$ are sampled using the uniform Latin Hypercube, with $\Theta^{b^\Delta_{ij}}\in[-7,7]^{18}$. Each 
   $\Theta^{b^\Delta_{ij}}$ vector corresponds to a candidate model, for which an OpenFOAM simulation of the PH$_{10595}$ case is carried out. With the chosen range, approximately $94\%$ of the sampled models prevents solution convergence. 
	For the remaining simulations, we extract the anisotropy tensor $b_{ij}$ at each mesh point and we examine the location of its eigenvalues with respect to the barycentric map. We observe that $84\%$ of the robust models yield realizable states at all mesh points. For these models, we compute the mean, the standard deviation and the minimum and maximum values for each component of $\Theta^{b^\Delta_{ij}}_l$. These quantities are shown in figure \ref{fig:realizability}, which shows that the coefficients yielding realizable states are approximately equal to 0 in the mean and have a variance of approximately 4. For the study conducted here, only 300 random samples have been considered. 
	Even though this number is not sufficient for the statistical quantities to achieve convergence, the results are in good accordance with the previously discussed \emph{a priori} realizability study, since each sample generates a Reynolds stress field and realizability can be checked at each mesh point.
	
	We then performed the same study with 300 random vectors $\Theta^{b^\Delta_{ij}}$ sampled in the range $[-2,2]^{18}$. More than 50 \% of the samples led to convergence of the simulation and approximately 70\% of these models yielded to fully realizable Reynolds stress fields. 
	We then optimize the model in the narrower parameter range $[-2,2]$, and check \emph{a posteriori} that the algorithm selects a realizable optimum.  
The computations can be run in parallel and some of them can be re-used to initialize the CORS algorithm at the subsequent optimization step.

We point out that the CFD-based realizability study can be skipped, since unrealizable models tend to be numerically unstable and in any case less accurate, and tend to be naturally discarded by the optimizer. In this case, arbitrary search ranges must be assigned by the user to the optimizer, and the realizability of the learned models is only assessed \emph{a posteriori}.
Based on the present \emph{a priori} and \emph{a posteriori} studies, we recommend using ranges of $[-2,2]$ or less.

	\subsubsection{Regularization terms}
	The regularization terms of the optimization problem (\ref{eq:CFD-driven_optimization_problem}) have to be specified. 
	Unlike the frozen-training SpaRTA, the CFD-driven training involves a nonlinear dependency of the cost function on the model parameters, due to the nonlinearity of the RANS equations.
	Sparsity-promoting regularization of nonlinear parameters would require an approximate computation of the cost function derivatives with respect to the parameters or function linearization \cite{kommenda2020parameter,pmlr-v48-yangc16}. 
	This can be efficiently done for CFD using an adjoint solver, which is not always promptly available. 
	For this reason, we choose to simply add a LASSO regularization, and we examine its effect on the complexity and generalizability of the resulting models.
The optimization problem of equation (\ref{eq:CFD-driven_optimization_problem}) is reformulated as:
\begin{equation}
{\Theta} =   \underset{\hat{\Theta}}{\operatorname{argmin}}      \| U - U_{RANS}(\hat{\Theta}) \|^2_2 + \lambda \| \hat{\Theta} \|_1 ,
\label{eq:optimization_CFD_driven_regularise}
\end{equation}  
where $\lambda$ is the regularization weight.
Since its optimal value is not known a priori, the optimization problem given in equation \ref{eq:optimization_CFD_driven_regularise} must be solved for $N$ values $\lambda_i$ with $1 \leq i \leq N$.
In particular in this work we have selected $\lambda=10^{-1},10^{-2},10^{-3},10^{-4},10^{-5},10^{-6}$.
This results in six alternative models for each test case of section 3.
Cross-validation \cite{bishop2006pattern} is then used to select the best-performing model across all validation data sets.
The performance of each candidate is arbitrarily evaluated as the mean squared error with respect to the high fidelity data. Other choices are possible.

\section{Numerical results}
\label{sec:numerical_results}

\subsection{Model discovery}
In the following, the CFD-driven SpaRTA is applied to the flow configurations of Section \ref{sec:datasets}.
Two series of numerical experiments are carried out, using different training data.

\begin{figure}[H]
	\centering
	\includegraphics[width=1.\linewidth]{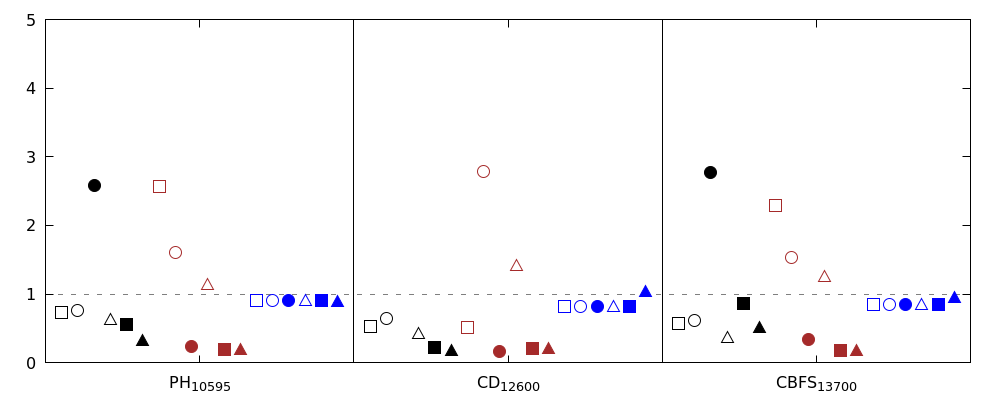} 
	\caption{ Cross-validation errors based on velocity profiles ($\overline{u_1}$, $\overline{u_2}$) for models trained on Reynolds-stress data. The errors are normalized with the mean-squared error of the baseline k-$\omega$ SST model. Black : models trained on PH$_{10595}$. \textcolor{red}{Red} : models trained on CD$_{12600}$. \textcolor{blue}{Blue} : models trained on CBFS$_{13700}$. $\Box$ : \textcolor{red}{$\lambda=10^{-1}$}. $\circ$ : \textcolor{red}{$\lambda=10^{-2}$}. $\bullet$ : \textcolor{red}{$\lambda=10^{-3}$}. $\bigtriangleup$ : \textcolor{red}{$\lambda=10^{-4}$}. $\blacksquare$ : \textcolor{red}{$\lambda=10^{-5}$}. $\blacktriangle$ : \textcolor{red}{$\lambda=10^{-6}$}.}
	\label{fig:convergence_apprentissage_tau}
\end{figure}

\begin{figure}[H]
	\centering
	\includegraphics[width=1.\linewidth]{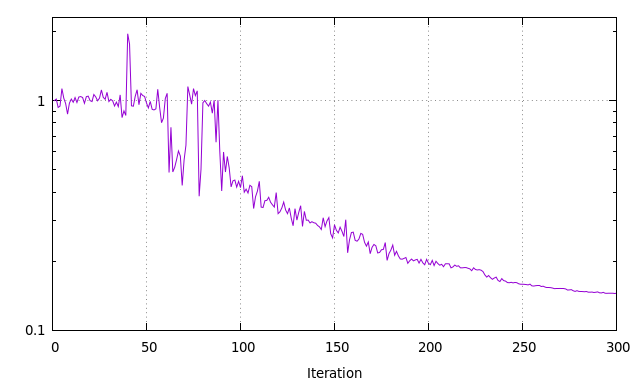} 
	\caption{Convergence of the optimization for Model 1 : value of the cost function with respect to the iteration.}
	\label{fig:convergence_history_tau}
\end{figure}

In a first series of experiments,  high-fidelity data for the Reynolds stresses are used to train the model, as in the frozen-training SpaRTA algorithm. 
However, in this case no additional data for the velocity field or the transported turbulent quantities is needed and the $k$-corrective frozen-RANS procedure is not applied.
More precisely, the training data are vertical profiles of $\tau_{11}$, $\tau_{22}$ and $\tau_{12}$. 
Profiles at $x/h=(0.05, 0.5,  1.0, 2.0, 3.0, 4.0, 5.0, 6.0, 7.0, 8.0 )$ are used for the  PH$_{10595}$ case;
for the CD$_{12600}$ we use profiles at $x/H=(5.5,6.5,7.5,8.5,9.5,10.5,11.5,12)$; finally, the profiles at $x/h=(0.1,1.5,3.0,4.5,6.0,7.5)$ are considered for case CBFS$_{13700}$. 
The amount of data used for the training is very sparse compared to the number of degrees of freedom used to solve the CFD problem: they represent 8.3\%, 5.7\% and 4.28\% of the total number of mesh points, used respectively for  PH$_{10595}$, CD$_{12600}$ and  CBFS$_{13700}$.
Six values of the regularization weight $\lambda$ are considered, as discussed previously.
For each of the six values of $\lambda$, the CORS algorithm has been run using $N=300$ function evaluations, and setting $n=39$ and $m=261$.
The model training step leads to the discovery of $3\times 6$ alternative models that are submitted to cross-validation: models trained for a flow configuration are applied to the two other cases. A total of 5400 RANS simulations (using 12 cores each on a standard workstation) have therefore been performed for the model discovery and the cross-validation. The overall cost of the learning procedure corresponds to approximately 16500 core-hours. 
This is obviously much more expensive than the  frozen-training procedure but it remains affordable without use of intensive computational facilities.
Mean squared errors on velocities, normalized by the mean-squared error of the baseline $k-\omega$ SST model, are reported in figure \ref{fig:convergence_apprentissage_tau}. Most models show an improvement over the baseline. Models trained on CBFS$_{13700}$ produce errors of the same order than the baseline model. Models trained on CD$_{12600}$ with low values of $\lambda$ produce the smallest errors averaged on the validation set.
The best-performing model, identified as the one having the lowest average error across the validation set, 
is trained on CD$_{12600}$ with $\lambda=10^{-5}$, and is referred-to as Model 1 in the following. The convergence history of the cost function residual for this model is shown in figure \ref{fig:convergence_history_tau}. We see that a satisfactory convergence of the optimization is reached. The model exhibits only a few terms that are bigger than $10^{-1}$ in magnitude.  The other terms, are mostly $O(10^{-2})$ or lower, and are neglected in the selected formulation of the model. 

The resulting mathematical expressions for the two correction tensors involve 4 out of the 36 candidate functions left after parameter space reduction:

\begin{equation}
 b_{ij}^\Delta =   -1.47 \times 10^{-1} I_1^2  T_{ij}^{(1)}  -2.6791 \times 10^{-1}   T_{ij}^{(2)} \\
\label{eq:M_CD_lambda=10-5_b_Delta}
\end{equation}

\begin{equation}
\begin{array}{l l}
b_{ij}^R = 4.6018\times  10^{-1}  T_{ij}^{(1)} -1.6779 \times 10^{-1}  T_{ij}^{(3)}
\end{array}
\label{eq:M_CD_lambda=10-5_b_R}
\end{equation}

\begin{figure}[H]
	\centering
	\includegraphics[width=1.0\linewidth]{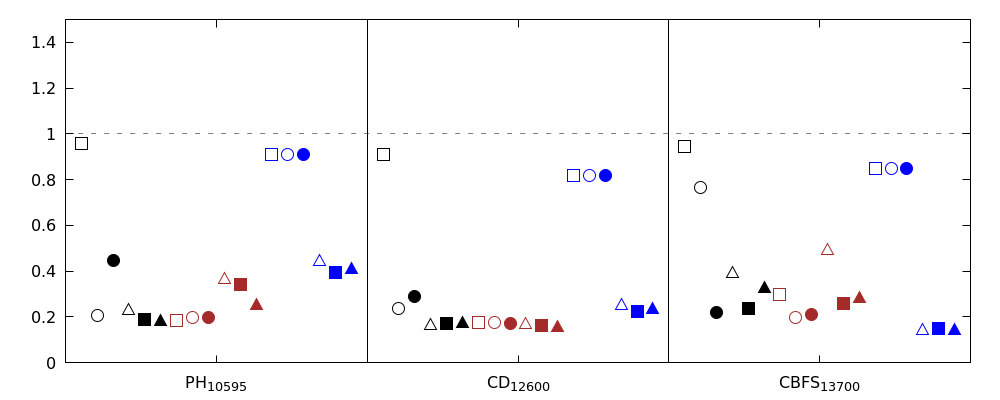} 
	\caption{Cross-validation errors based on velocity profiles ($\overline{u_1}$, $\overline{u_2}$) for models trained on velocity and skin friction data. The errors are normalized with the mean-squared error of the baseline k-$\omega$ SST model. Black : models trained on PH$_{10595}$. \textcolor{red}{Red} : models trained on CD$_{12600}$. \textcolor{blue}{Blue} : models trained on CBFS$_{13700}$. $\Box$ : \textcolor{red}{$\lambda=10^{-1}$}. $\circ$ : \textcolor{red}{$\lambda=10^{-2}$}. $\bullet$ : \textcolor{red}{$\lambda=10^{-3}$}. $\bigtriangleup$ : \textcolor{red}{$\lambda=10^{-4}$}. $\blacksquare$ : \textcolor{red}{$\lambda=10^{-5}$}. $\blacktriangle$ : \textcolor{red}{$\lambda=10^{-6}$}.}
	\label{fig:convergence_apprentissage_U_Cf}
\end{figure}

\vspace{1cm}
In the second series of experiments, the method is trained against high fidelity velocity profiles and skin friction data to test its ability to discover models optimized for any target quantity of interest and to assess the sensitivity of the
symbolic identification procedure to the kind of training data.
Velocity profiles ($\overline{u_1}$ and $\overline{u_2}$) are extracted at the same locations of the Reynolds stress profiles, whereas the
skin friction data are extracted along the bottom walls in all cases.
In this setting the data represent 9.1\%, 6.7\% and 4.95\% of the total number of mesh points for  PH$_{10595}$, CD$_{12600}$ and  CBFS$_{13700}$, respectively.
Cross validation errors for the 18 discovered models are reported in figure \ref{fig:convergence_apprentissage_U_Cf}. The target variable (the longitudinal velocity) being the same as the training one, the validation errors are lower.

The best-performing model is the one learned on CD$_{12600}$ with $\lambda=10^{-2}$ and it is hereafter named Model 2.
The convergence history of the cost function residual for training of Model 2 is shown in figure \ref{fig:convergence_history_U_Cf}. Again, a satisfactory convergence of the optimization is reached. 
As for Model 1, Model 2 is sparse and the retained mathematical expression with coefficients $O(10^{-1})$ or higher is:

\begin{figure}[H]
	\centering
	\includegraphics[width=1.\linewidth]{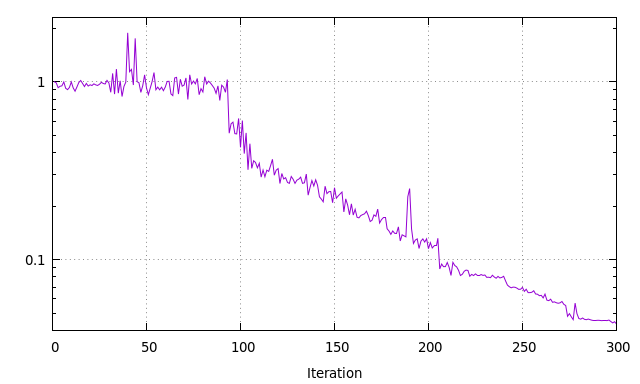} 
	\caption{Convergence of the optimization for Model 2 : value of the cost function with respect to the iteration.}
	\label{fig:convergence_history_U_Cf}
\end{figure} 

\begin{equation}
\begin{array}{l l}
 b_{ij}^\Delta =  -2.8356 \times 10^{-1}  T_{ij}^{(1)}   -1.4738 \times 10^{-1} I_2^2  T_{ij}^{(2)} \\
\end{array}
\label{eq:M_CD_lambda=10-2_b_Delta}
\end{equation}

\begin{equation}
\begin{array}{l l}
b_{ij}^R = &  ( 1.0375 \times 10^{-1} I_2 -2.8833 \times 10^{-1} I_1I_2) T_{ij}^{(3)}
\end{array}
\label{eq:M_CD_lambda=10-2_b_R}
\end{equation}

For both models we checked a posteriori that the realizability conditions are satisfied throughout the flow domain for all cases.

In the following, a detailed assessment of the two learned models against the high-fidelity data is carried out for a number of flow quantities, and the results are compared to those of the $k-\omega$ SST model.
For comparison, we also include the results of the BSL-EARSM model of Menter et al. \cite{menter2012explicit} and those of a frozen-training SpaRTA model reported in Ref. \cite{schmelzer2020machine}, given in the following:
\begin{equation}
\begin{array}{l l}
b_{ij}^\Delta &=0
\end{array}
\label{eq:M_Sparta1_b_Delta}
\end{equation}

\begin{equation}
\begin{array}{l l}
b^R_{ij} &=3.9\times 10^{-1} T_{ij}^{(1)}
\end{array}
\label{eq:Sparta1}
\end{equation}
and that corresponds to a simple correction of the production terms in the $k-\omega$ transport equations.

Finally, the learned models are applied to an unseen flow case not included in the training and testing datasets, namely the periodic hill flow at $Re=37000$, and the results are compared to the reference data and to the models mentioned in the above.

\subsection{Results for flow configurations in the training and validation sets}
\begin{figure}[H]	
	\centering
	\begin{subfigure}[b]{1.0\textwidth}
		\centering
		\includegraphics[width=\textwidth]{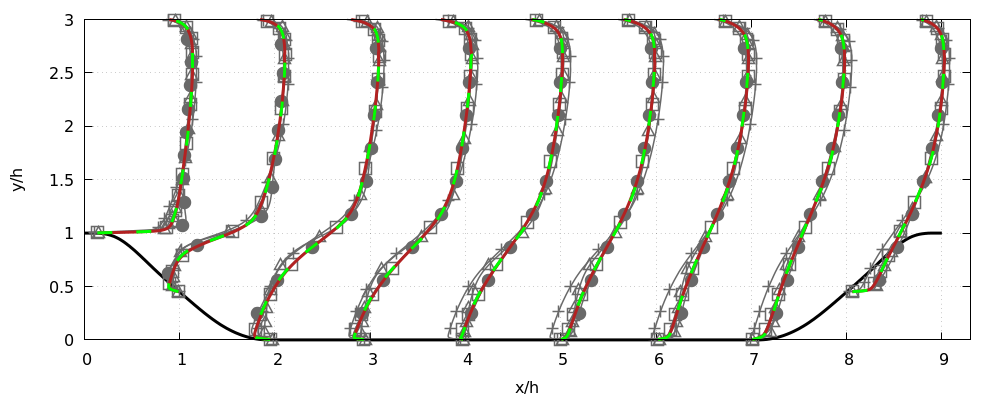}
		\caption{PH$_{10595}$}
		\label{fig:y equals x}
	\end{subfigure}
	\vfill
	\begin{subfigure}[b]{1.0\textwidth}
		\centering
		\includegraphics[width=\textwidth]{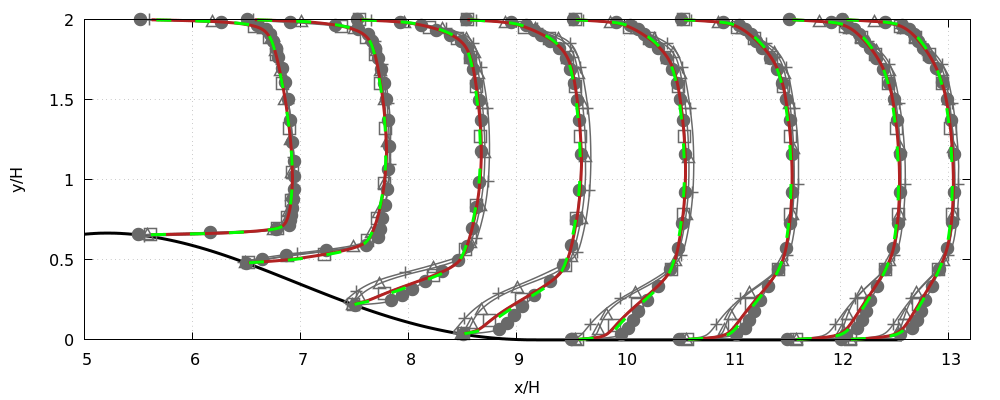}
		\caption{CD$_{12600}$}
		\label{fig:three sin x}
	\end{subfigure}
	\vfill
	\begin{subfigure}[b]{1.0\textwidth}
		\centering
		\includegraphics[width=\textwidth]{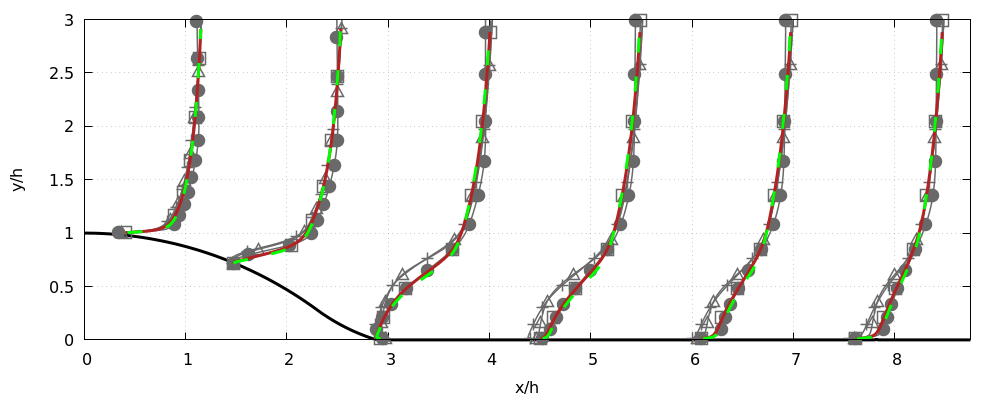}
		\caption{CBFS$_{13700}$}
		\label{fig:five over x}
	\end{subfigure}
	\caption{Assessment of various CFD-driven, frozen-training  and standard turbulence models for flows in the training and test set: streamwise velocity profiles ($\overline{u_1}/U_b+x/h$). k-$\omega$-SST: (\protect\cross), High fidelity data : (\protect\circlee), Model 1 : (\protect\redline), Model 2 : (\protect\greendashedline), frozen-training:  
		 (\protect\rectangle), BSL-EARSM : (\protect\trianglee). }
	\label{fig:vitesses_apprentissage_tau}
\end{figure}

\begin{figure}[H]	
	\centering
	\begin{subfigure}[b]{1.0\textwidth}
		\centering
		\includegraphics[width=\textwidth]{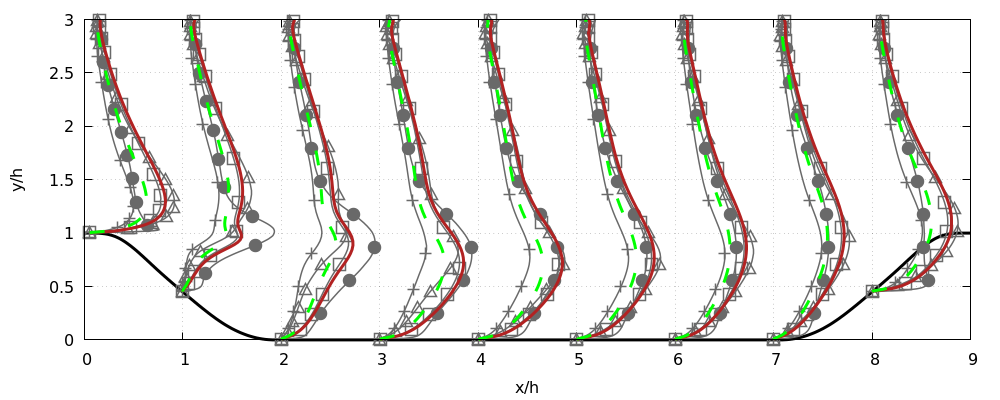}
		\caption{ PH$_{10595}$}
		\label{fig:y equals 1}
	\end{subfigure}
	\vfill
	\begin{subfigure}[b]{1.0\textwidth}
		\centering
		\includegraphics[width=\textwidth]{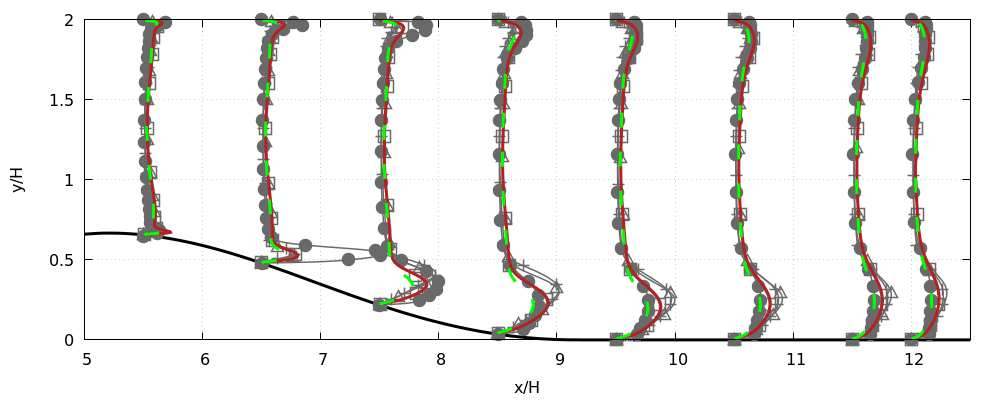}
		\caption{CD$_{12600}$}
		\label{fig:three sin 1}
	\end{subfigure}
	\vfill
	\begin{subfigure}[b]{1.0\textwidth}
		\centering
		\includegraphics[width=\textwidth]{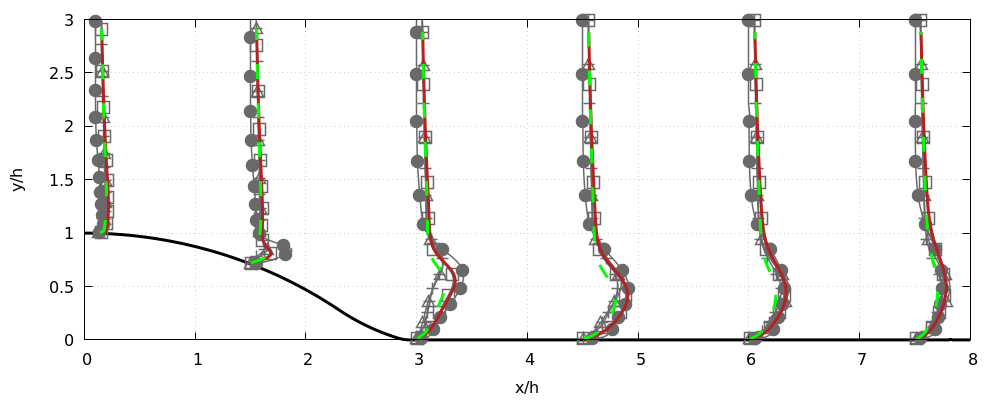}
		\caption{CBFS$_{13700}$}
		\label{fig:five over 1}
	\end{subfigure}
	\caption{Assessment of various CFD-driven, frozen-training  and standard turbulence models for flows in the training and test set: turbulent kinetic energy profiles ($12k/U_b^2+x/h$). k-$\omega$-SST: (\protect\cross), High fidelity data : (\protect\circlee), Model 1 : (\protect\redline), Model 2 : (\protect\greendashedline), frozen-training:  
		 (\protect\rectangle), BSL-EARSM : (\protect\trianglee).}
	\label{fig:k_apprentissage_tau}
\end{figure}

\begin{figure}[H]	
	\centering
	\begin{subfigure}[b]{1.0\textwidth}
		\centering
		\includegraphics[width=\textwidth]{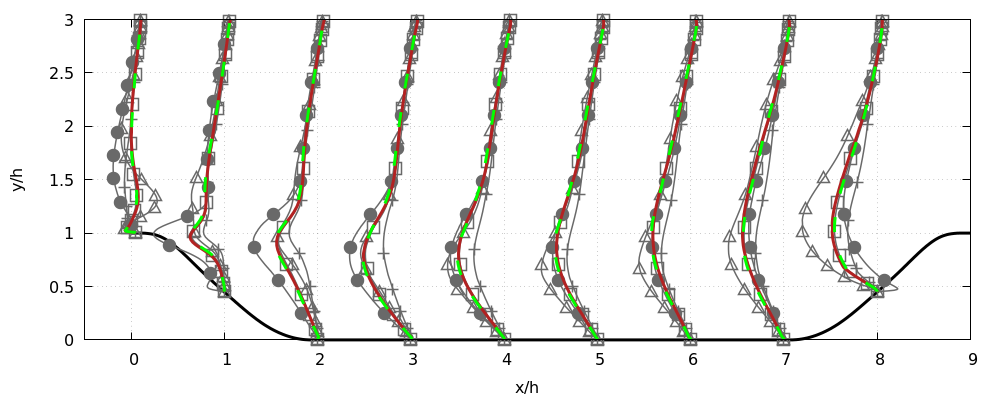}
		\caption{PH$_{10595}$}
		\label{fig:y equals 2}
	\end{subfigure}
	\vfill
	\begin{subfigure}[b]{1.0\textwidth}
		\centering
		\includegraphics[width=\textwidth]{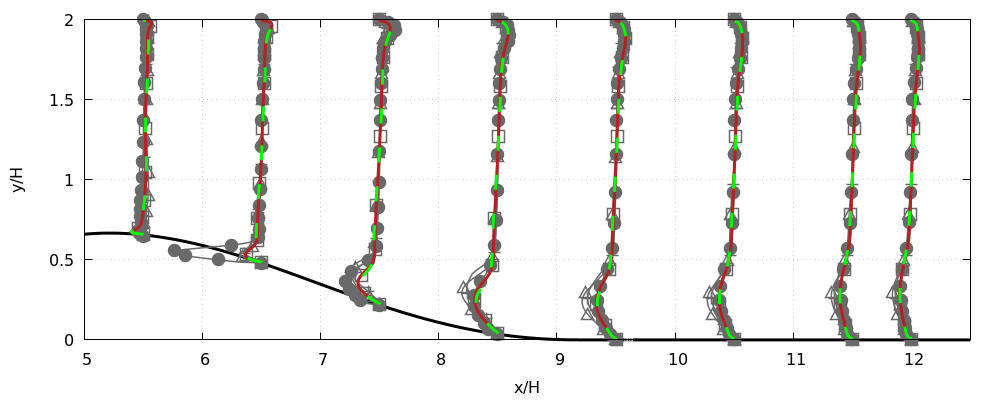}
		\caption{CD$_{12600}$}
		\label{fig:three sin 2}
	\end{subfigure}
	\vfill
	\begin{subfigure}[b]{1.0\textwidth}
		\centering
		\includegraphics[width=\textwidth]{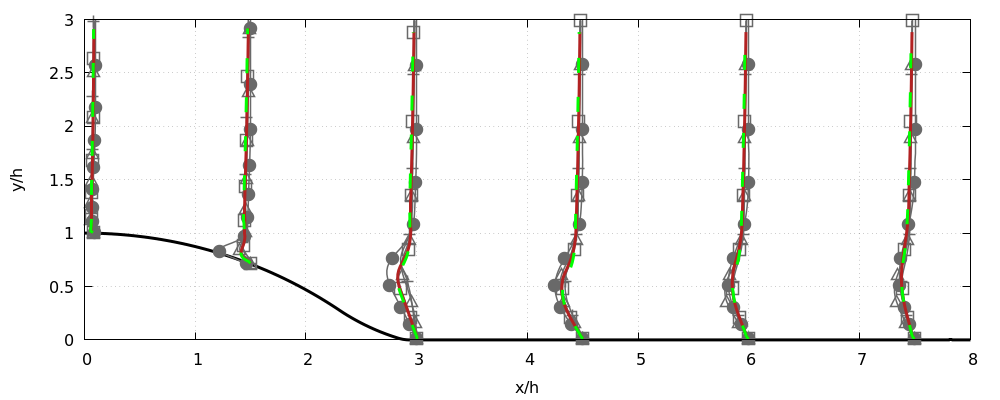}
		\caption{CBFS$_{13700}$}
		\label{fig:five over 2}
	\end{subfigure}
	\caption{Assessment of various CFD-driven, frozen-training  and standard turbulence models for flows in the training and test set:  Reynolds shear stress profiles ($20\tau_{12}/U_b^2+x/h$). k-$\omega$-SST: (\protect\cross), High fidelity data : (\protect\circlee), Model 1 : (\protect\redline), Model 2 : (\protect\greendashedline), frozen-training:  
		 (\protect\rectangle), BSL-EARSM : (\protect\trianglee).}
	\label{fig:tauxy_apprentissage_tau}
\end{figure}

\begin{figure}[H]	
	\centering
	\begin{subfigure}[b]{1.0\textwidth}
		\centering
		\includegraphics[width=\textwidth]{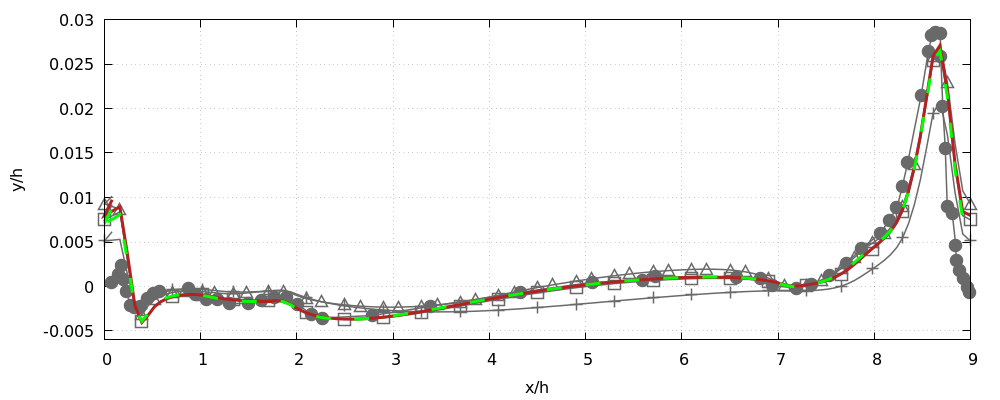}
		\caption{PH$_{10595}$}
		\label{fig:Cfy equals 2}
	\end{subfigure}
	\vfill
	\begin{subfigure}[b]{1.0\textwidth}
		\centering
		\includegraphics[width=\textwidth]{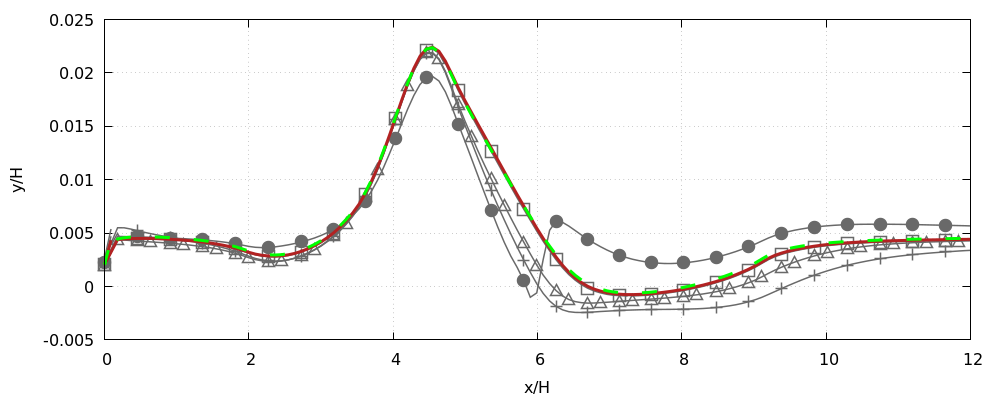}
		\caption{CD$_{12600}$}
		\label{fig:Cfthree sin 2}
	\end{subfigure}
	\vfill
	\begin{subfigure}[b]{1.0\textwidth}
		\centering
		\includegraphics[width=\textwidth]{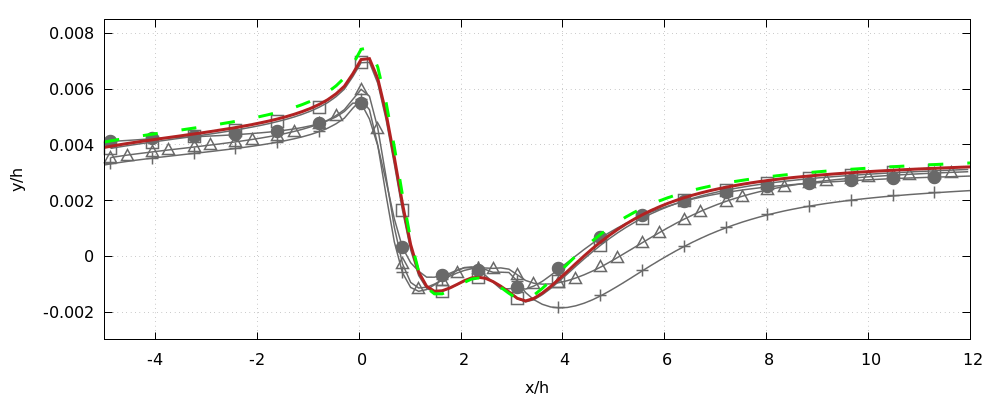}
		\caption{CBFS$_{13700}$}
		\label{fig:Cffive over 2}
	\end{subfigure}
	\caption{Assessment of various CFD-driven, frozen-training and standard turbulence models for flows in the training and test set: skin friction distributions along the bottom wall.  k-$\omega$-SST: (\protect\cross), High fidelity data : (\protect\circlee), Model 1 : (\protect\redline), Model 2 : (\protect\greendashedline), frozen-training:  
		 (\protect\rectangle), BSL-EARSM : (\protect\trianglee).}
	\label{fig:Cf_apprentissage_tau}
\end{figure}

The predictions of the discovered models Model 1 and Model 2 for the 3 test cases are presented in figures \ref{fig:vitesses_apprentissage_tau}, \ref{fig:k_apprentissage_tau}, \ref{fig:tauxy_apprentissage_tau} and \ref{fig:Cf_apprentissage_tau} for the streamwise velocity $\overline{u_1}$, the turbulent kinetic energy $k$, the Reynolds shear stress $\tau_{12}$ and the skin friction coefficient $C_f$, respectively. Their results are compared with those of the baseline k-$\omega$ SST model, the frozen-training SpaRTA model, the BSL-EARSM model and the high-fidelity data. 
For a more quantitative evaluation we report in table \ref{table:errors_model_tau} the mean squared errors on the longitudinal and wall-normal velocities, turbulent kinetic energy and Reynolds shear stress ($\tau_{12}$) profiles, as well as for the skin friction coefficient $C_f$, respectively. In all cases, the errors are normalized with the mean-squared error of the baseline k-$\omega$ SST model.  
\begin{table}[H]
\caption{Mean-squared errors of predicted velocity, turbulent kinetic energy and  Reynolds shear-stress profiles and skin friction distribution with respect to high-fidelity data for various models. The errors are normalized by the mean-squared error of the baseline k-$\omega$ SST model.}
	\centering
	\subcaption*{Velocity ($(\overline{u}_1,\overline{u}_2)$}
	\begin{tabular}{|l|c|c|c|}
		\hline
		& PH$_{10595}$  & CD$_{12600}$  & CBFS$_{13700}$   \\\hline
		Model 1  & 0.1901 & 0.1993 & 0.1687 \\
		\hline
		Model 2  & 0.1932  &  0.1767 & 0.2012 \\
		\hline
	    Frozen-training   & 0.2067    &  0.2067 &  0.1711\\
		\hline
	    BSL-EARSM  &  0.3207  &  0.5999  &   0.7762\\
		\hline
	\end{tabular}
%
\vspace{0.2cm}
	\centering
	\subcaption*{Turbulent kinetic energy $k$}
	\begin{tabular}{|l|c|c|c|}
		\hline
		& PH$_{10595}$  & CD$_{12600}$  & CBFS$_{13700}$   \\\hline
		Model 1  & 0.3677  & 0.7279 & 0.6136   \\
		\hline
		Model 2  & 0.4740  & 0.9510 & 0.9889 \\
		\hline
	         Frozen-training  & 0.3478   &  0.7472  & 0.6135 \\
		\hline
	    BSL-EARSM  &    0.6881     &           0.9821       &     1.0065  \\
		\hline
	\end{tabular}
%
\vspace{0.2cm}
	\centering
	\subcaption*{Reynolds shear stress $\tau_{12}$}
	\begin{tabular}{|l|c|c|c|}
		\hline
		& PH$_{10595}$  & CD$_{12600}$  & CBFS$_{13700}$   \\\hline
		Model 1  & 0.5485 &  0.9204 & 0.721    \\
		\hline
				Model 2  & 0.519 &0.9545 &0.834 \\
		\hline
			    Frozen-training  & 0.5535   &  0.9394 & 0.7162 \\
		\hline
			    BSL-EARSM  &  1.2229  &   0.9524   &        0.8061  \\
		\hline
	\end{tabular}
%
\vspace{0.2cm}
	\centering
	\subcaption*{Skin friction coefficient $C_f$}
	\begin{tabular}{|l|c|c|c|}
		\hline
		& PH$_{10595}$  & CD$_{12600}$  & CBFS$_{13700}$   \\\hline
		{Model 1}  & {0.6828}  & {0.6004} & {0.3194}   \\
		\hline
				{Model 2}  &  0.5888  & {0.5548} & {0.5836}  \\
		\hline
			   Frozen-training  &  0.6427 &   0.5958 & 0.3125 \\
		\hline
			    BSL-EARSM  &   0.8443  &  0.6383  &  0.2203  \\
		\hline
	\end{tabular}
	\label{table:errors_model_tau}
\end{table}

Both discovered models drastically improve velocity when compared to the $k-\omega$-SST and the BSL-EARSM models, particularly within the separated regions (see figure  \ref{fig:vitesses_apprentissage_tau}) for the 3 flow cases. 
Furthermore, their accuracy is similar to the frozen-training SpaRTA. Model 2, learned using high fidelity velocity profiles and skin friction training data, provides a lower error in velocity  than Model 1, learned using high fidelity Reynolds stress profiles as training data. 

The discovered models improve the prediction of $k$ profiles over the LEVM $k-\omega$ SST for the 3 learning flow configurations. As expected, Model 1 yields better prediction of turbulent kinetic energy than Model 2 for all cases. Nonetheless, Model 2 overall improves the prediction of $k$ with respect to the baseline. Of note,  the accuracy of Model 1 and the {\color{blue}frozen-training} model (also trained Reynolds stresses ) 
in predicting $k$ is very similar for the 3 flow cases.

The discovered models also improve the prediction of $\tau_{12}$ profiles over the LEVM (see figure \ref{fig:tauxy_apprentissage_tau} and table \ref{table:errors_model_tau}), Model 1 having better prediction capabilities than Model 2 and its accuracy is similar to the frozen-training model.

The skin friction results are reported in figure \ref{fig:Cf_apprentissage_tau} and table \ref{table:errors_model_tau}). As expected, Model 2, trained on the skin friction, yields better predictions than Model 1 and the frozen-training model. Both models improve the prediction of the reattachment point location for cases PH$_{10595}$ and CBFS$_{13700}$ with respect to the predictions of the $k-\omega$ SST model, as shown in table \ref{table:reattachement_position_tau}.  However, neither of the two models captures the small recirculation bubble predicted by the DNS of case CD$_{12600}$. The predictions of the separation and reattachment point locations for case CD$_{12600}$ are therefore not shown in table \ref{table:reattachement_position_tau}. 

\begin{table}[H]
\caption{Predicted separation and reattachement points ($x/h$) for various models.}
	\label{table:reattachement_position_tau}

	\centering
	\subcaption*{Separation points}\vspace{-0.1cm}
	\begin{tabular}{|l|c|c|}
		\hline
    	& PH$_{10595}$   & CBFS$_{13700}$   \\\hline
        LES  & 0.2154  &  0.9429 \\\hline
        k-$\omega$ SST  & 0.2696   &  0.7655 \\\hline
		Model 1  & {0.278}  & {1.0895}  \\ \hline
    	Model 2  & {0.278}  & {1.0895}   \\\hline
    	Frozen-training   & 0.2845 &    1.0503  \\\hline
    	BSL-EARSM  &    0.2817       &       0.8181   \\\hline
	\end{tabular}
        \centering
        \vspace{0.3cm}
	\subcaption*{Reattachment points}\vspace{-0.1cm}
	\begin{tabular}{|l|c|c|}
		\hline
		& PH$_{10595}$   & CBFS$_{13700}$   \\\hline
		LES  & 4.7595  &  4.2409\\\hline
		k-$\omega$ SST  & 7.6418   &  6.0296 \\\hline
		{Model 1}  & {4.860}  &  {4.3334}  \\
		\hline
	    {Model 2}  & {4.860}  &  {4.1694}  \\
		\hline
			    Frozen-training   &  5.0115 &    4.4754 \\
		\hline
			    BSL-EARSM  &   4.5613    &         5.1107 \\
		\hline
	\end{tabular}
	\end{table}

\subsection{Model learning from full-field high fidelity data}
In this Section, we investigate the effect of the amount of data required to train the model. For that reason we retrained Model 1 using the full high-fidelity fields for the Reynolds stresses, interpolated on the RANS computational grid.

The resulting learned model is hereafter referred-to as Model 1FF (FF : Full Field). The correction tensors involve only 1 out of the 36 candidate functions:
\begin{equation}
b_{ij}^\Delta = 0,
\label{eq:M_CD_lambda=10-5_fullfield_b_Delta}
\end{equation}
\begin{equation}
b_{ij}^R =4.0120\times 10^{-1} T_{ij}^{(1)}.
\label{eq:M_CD_lambda=10-5_fullfield_b_R}
\end{equation}

We report in table \ref{table:errors_model_tau_FF} the mean squared errors on the longitudinal and wall-normal velocities, turbulent kinetic energy and Reynolds shear stress, and skin friction coefficient. We note that drastically increasing the amount of data does not improve solution accuracy.

Interestingly, the formulation of Model 1FF is very close to the formulation of model M1 of Ref. \cite{schmelzer2020machine} (see equations (\ref{eq:M_Sparta1_b_Delta}) and (\ref{eq:Sparta1})) trained on the same data. Moreover Model 1, Model 1FF and the full-field frozen-training model M1 have similar coefficients of the dominant term ($4.6018\times 10^{-1}$, $4.0120\times 10^{-1}$ and $3.9\times 10^{-1}$, respectively). A possible interpretation is as follows: although more accurate than the baseline LEVM, a generalized eddy viscosity model remains an imperfect model that cannot match the data even with the best possible set of parameters (i.e. the mean squared error cannot be reduced below a given threshold). The algorithm compensate this effect by minimizing the second term in the cost function, leading to a sparser model.

\begin{table}[H]
\caption{Mean-squared errors of predicted velocity, turbulent kinetic energy and  Reynolds shear-stress profiles and skin friction distribution with respect to high-fidelity data for Model 1FF. The errors are normalized by the mean-squared error of the baseline k-$\omega$ SST model.}\label{table:errors_model_tau_FF}
	\centering
	\begin{tabular}{|l|c|c|c|c|}
		\hline
		&$\overline{u}_1,\overline{u}_2$    & $k$ & $\tau_{12}$ & $C_f$ \\\hline
		$PH_{10595}$  & 0.2029 & 0.3488 & 0.5524 & 0.6474 \\\hline 
		$CD_{12600}$  & 0.2048  & 0.7440  & 0.9388 & 0.5956
		\\\hline
		 $CBFS_{13700}$  &  0.1694  & 0.6144  & 0.7209  & 0.3232 \\\hline
	\end{tabular}
\end{table}

\vspace{1cm}
To better discriminate among the learned models, in the next Section we evaluate their extrapolation performance to a flow configuration outside the training and test sets.
\subsection{Extrapolation to an unseen flow}
\begin{figure}[H]	
	\centering
	\includegraphics[width=\textwidth]{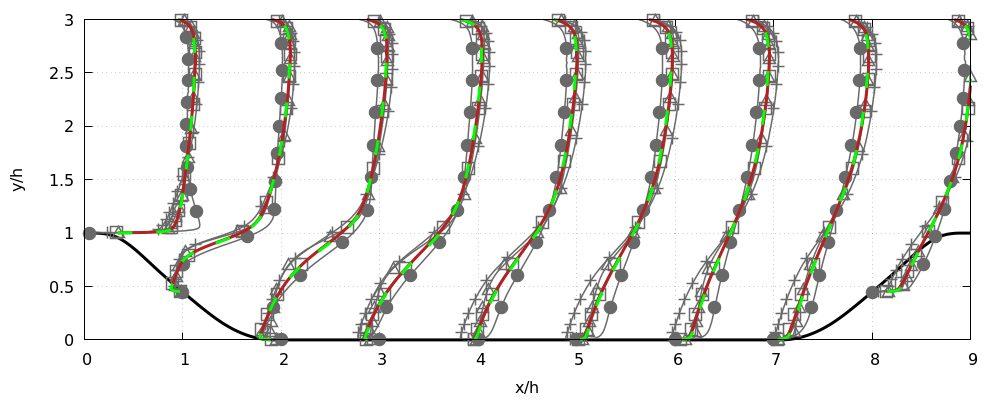}
	\caption{Longitudinal velocity profiles normalized with bulk velocity : $\overline{u_1}/U_b+x$ for PH$_{37000}$. k-$\omega$-SST: (\protect\cross), High fidelity data : (\protect\circlee), Model 1 : (\protect\redline), Model 2 : (\protect\greendashedline), frozen-training:  
		 (\protect\rectangle), BSL-EARSM : (\protect\trianglee).}
	\label{fig:PH_Re_37000_U_TAU}
\end{figure}

\begin{figure}[H]	
	\centering
	\includegraphics[width=\textwidth]{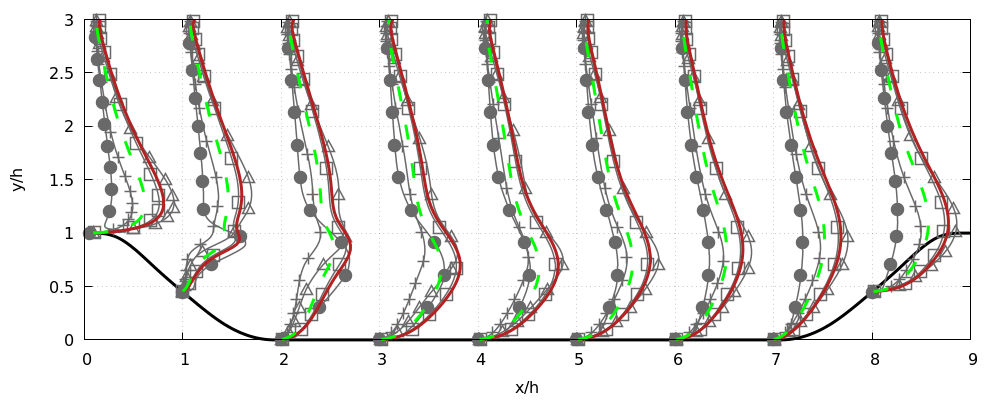}
	\caption{Turbulent kinetic energy profiles normalized with bulk velocity : $12k/U_b^2+x$ for PH$_{37000}$. k-$\omega$-SST: (\protect\cross), High fidelity data : (\protect\circlee), Model 1 : (\protect\redline), Model 2 : (\protect\greendashedline), frozen-training:  
		 (\protect\rectangle), BSL-EARSM : (\protect\trianglee).}
	\label{fig:PH_Re_37000_k_TAU}
\end{figure}

\begin{figure}[H]	
	\centering
	\includegraphics[width=\textwidth]{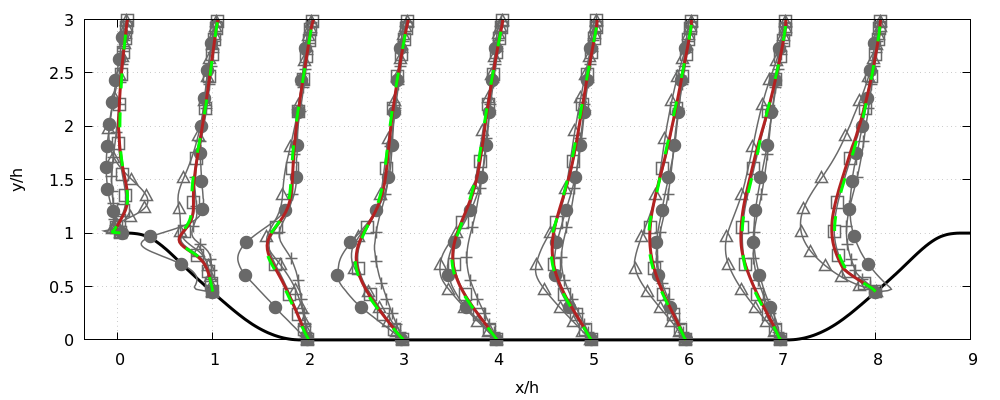}
	\caption{Reynolds shear stress profiles normalized with bulk velocity : $20\tau_{12}/U_b^2+x$ for PH$_{37000}$. k-$\omega$-SST: (\protect\cross), High fidelity data : (\protect\circlee), Model 1 : (\protect\redline), Model 2 : (\protect\greendashedline), frozen-training:  
		 (\protect\rectangle), BSL-EARSM : (\protect\trianglee).}
	\label{fig:PH_Re_37000_tauxy_TAU}
\end{figure}

\begin{figure}[H]	
	\centering
	\includegraphics[width=\textwidth]{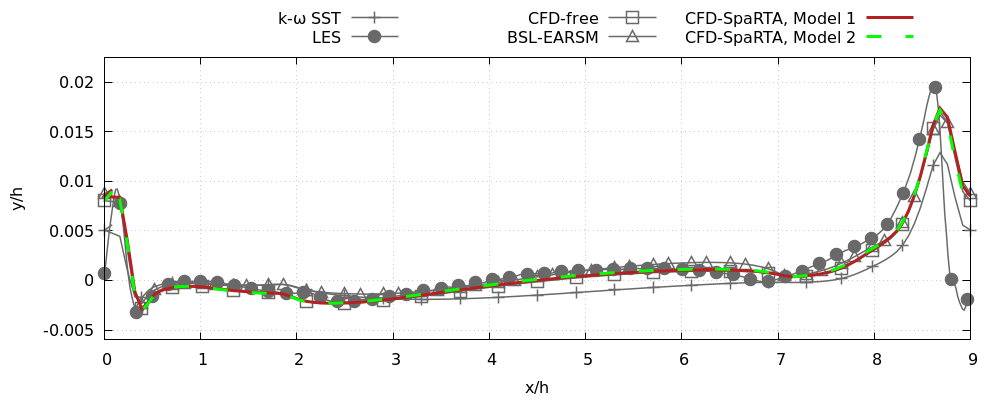}
	\caption{Skin friction distribution $C_f$ for PH$_{37000}$.  k-$\omega$-SST: (\protect\cross), High fidelity data : (\protect\circlee), Model 1 : (\protect\redline), Model 2 : (\protect\greendashedline), frozen-training:  
		 (\protect\rectangle), BSL-EARSM : (\protect\trianglee).}
	\label{fig:PH_Re_37000_Cf_TAU}
\end{figure}

To test the ability of the CFD-driven SpaRTA to serve as a predictive model for 2D turbulent flow with separations, the two CFD-driven models learned for cases PH$_{10595}$, CD$_{12600}$ and CBFS$_{13700}$ are applied to the 2D periodic hill flow case at Re=37000 (noted PH$_{37000}$).
Model 1 and Model 2 have been learned on CD$_{12600}$, therefore, for both models, PH$_{37000}$ represents an extrapolation to a higher Reynolds number and a different geometry, even if the flow still belongs to the class of 2D incompressible separated flow in a variable section channel.
The results for the streamwise velocity $\overline{u_1}$, the turbulent kinetic energy $k$, the Reynolds shear stress $\tau_{12}$ and the skin friction coefficient $C_f$ are compared to LES data from \cite{gloerfelt2019large} in figures \ref{fig:PH_Re_37000_U_TAU}, \ref{fig:PH_Re_37000_k_TAU}, \ref{fig:PH_Re_37000_tauxy_TAU} and \ref{fig:PH_Re_37000_Cf_TAU}, respectively. 
The results of the baseline LEVM, of the BSL-EARSM and of the frozen-training  SpaRTA model are also reported for comparison.
\begin{table}[H]
\caption{Mean-squared errors of predicted velocity, turbulent kinetic energy and  Reynolds shear-stress profiles and skin friction distribution with respect to high-fidelity data for various models applied to the PH case at Re=37000. 
The errors are normalized by the mean-squared error of the baseline k-$\omega$ SST model.}\label{table:errors_model_TAU_PH37000}
	\centering
	\begin{tabular}{|l|c|c|c|c|}
		\hline
		&$\overline{u}_1,\overline{u}_2$    & $k$ & $\tau_{12}$ & $C_f$ \\\hline
		{Model 1}  & {0.2926} & {3.0006} & {0.6803} & {0.9828} \\\hline 
		{Model 2}  & {0.2726}  & {1.1259}  & {0.6606} & {0.8911}
		\\\hline
		 Frozen-training  &   0.3280  & 2.5153   & 0.6841  & 0.9292 \\\hline
		BSL-EARSM &    0.4380  & 3.7959 & 1.4510 &  0.9912 \\
\hline
	\end{tabular}
\end{table}

The normalized mean squared errors with respect to the high-fidelity data from \cite{gloerfelt2019large} are reported in table \ref{table:errors_model_TAU_PH37000} and the predicted separation and reattachment points are given
in table \ref{table:reattachement_position_PH37000}.

Model 1 and Model 2 improve the predictions of all quantities of interest, except the turbulence kinetic energy, showing their ability to improve the accuracy of the baseline k-$\omega$ SST even at a higher Reynolds number and for a different geometry. Model 2 is more accurate than \textcolor{orange}{Model 1} and the {\color{blue}frozen-training} for each considered quantity of interest, while Model 1 yields overall the same accuracy as the frozen-training model.

BSL-EARSM provides poor predictions with respect to the learned models for PH$_{37000}$, degrading the accuracy of the k-$\omega$ SST model for 2 of the 4 quantities, one quantity being predicted with the same accuracy than the k-$\omega$ SST model.

These results show that the CFD-driven models robustly improve predictive accuracy of the k-$\omega$ SST model also for a 2D separated flow outside the training set and exhibit a comparable extrapolation performance than the frozen-training SpaRTA. The extrapolation performance of Model 2 is even better than the one of the k-$\omega$ SST model.

\begin{table}[H]
	\caption{Predicted separation and reattachement points ($x/h$) for various models.}
	\label{table:reattachement_position_PH37000}
	\centering
	\begin{tabular}{|l|c|c|}
		\hline
		& separation   & reattachement   \\\hline
		LES  &  0.2552  &  4.0170 \\\hline
		k-$\omega$ SST  & 0.2679   & 7.5479  \\\hline
		Model 1  & {0.2780}  & {4.4600} \\
		\hline
		{Model 2}  & {0.2780} & {4.3799}   \\
		\hline
		Frozen-training   &  0.2951  &   4.6303   \\
		\hline
		BSL-EARSM  &   0.2898      &    4.2333    \\
		\hline
	\end{tabular}
\end{table}

\section{Conclusions}
A CFD-driven counterpart of the SpaRTA method initially proposed in \cite{schmelzer2020machine} has been presented. The new method aims at discovering data-augmented EARSMs with improved performance for a given class of flows with respect to a baseline LEVM  through deterministic symbolic identification of terms from a predefined library of candidate functions.
As in SpaRTA, the discovered models use a generalized eddy viscosity formulation in conjunction with corrective terms for the production of the transported turbulent variables.
The CFD-driven approach is intended to overcome limitations of the frozen-training  model discovery in terms of numerical robustness of the learned models and to enable flexible use of any kind of data for model training.
Importantly, the CFD-driven approach does not necessarily require data for second-order turbulence statistics, namely, the Reynolds stresses, which are not always easily available, e.g., in experimental data sets.

In the present work we focused on correcting the $k-\omega$ SST model for a class of 2D turbulent flows with separations caused by an enlargement of the channel section. 
The methodology may however be adapted to other eddy viscosity models and different flow configurations. The present test cases include the well-known periodic hill flow, a converging-diverging channel and the flow behind a curved backward facing step, for which high-fidelity DNS or LES data are available in the literature. 
To show the ability of the CFD-driven method to train models with any kind of data, two sets of high-fidelity data have been considered. First, models are learned using data for Reynolds stress profiles at a few streamwise locations.
Second, we prove the flexibility of the proposed approach by training models against selected velocity profiles and skin friction data. 

The computational cost of CFD-driven model identification is obviously greater than frozen-training  approach. To speed up the learning process, we use a surrogate model for solving the optimization problem. By so doing, a complete model identification run, including model training and selection via cross-validation, requires approximately 5000 CFD solves, which remains affordable even on a standard workstation for 2D flow cases. 

The CFD-driven discovered models are sparse and numerically robust, thanks to the regularization constraints on model parameters. They satisfy realizability conditions throughout the computational domain.
For flow configurations in the training and validation sets, they exhibit improved accuracy not only over the baseline LEVM, but also over a general purpose EARSM model \cite{menter2012explicit}. 
Furthermore, their have an accuracy comparable to a data-driven model generated with the frozen-training SpaRTA algorithm.
The model learned from velocity and skin friction data is overall more accurate than the one trained on Reynolds stress data for the prediction of the velocity and skin friction. On the opposite, the model trained on the latter data is overall more accurate in predicting the turbulent kinetic energy and the Reynolds shear stress.

The effect of the amount of data required to train the model has then been investigated. Model 1 has been retrained using the full high-fidelity fields for the Reynolds stresses, interpolated on the RANS computational grid. We showed that drastically increasing the the amount of data does not improve solution accuracy, but leads to a sparser model, whose formulation is very close to the frozen-training SpaRTA M1 model of Ref. \cite{schmelzer2020machine}, trained on the same data.

The CFD-driven models have then been applied to a separated flow outside the
training and validation sets, characterized by a higher Reynolds number and a different geometry. 
For this configuration, the learned models provide significantly improved results for most flow quantities of interest, not only over the LEVM but also over the general purpose BLS-EARSM model. 
The CFD-driven models yield a comparable or better accuracy than the frozen-training  SpaRTA.

In conclusion, the CFD-driven SpaRTA represents a promising method to generate data-driven EARSMs. The computational overcost of the training algorithm is justified by the possibility of using any kind of available data, including incomplete data sets, as those potentially obtained from experiments. Furthermore, such models are by construction numerically robust. 


\section*{Code}
The code of the frozen-training SpaRTA has been made public and can be found in the following github repository : \href{https://github.com/shmlzr/general_earsm.git}{https://github.com/shmlzr/general\_earsm.git}.

\section{ Acknowledgements }

The present work was carried out in the frame of project PERTINENT (high-fidelity exPerimental and numERIcal study for daTa-drIven modeliNg of rEalistic turbuleNt separaTed flows), funded by Institut Carnot ARTS, France.

\bibliography{mybibfile}

\begin{thebibliography}{10}
\expandafter\ifx\csname url\endcsname\relax
  \def\url#1{\texttt{#1}}\fi
\expandafter\ifx\csname urlprefix\endcsname\relax\def\urlprefix{URL }\fi
\expandafter\ifx\csname href\endcsname\relax
  \def\href#1#2{#2} \def\path#1{#1}\fi

\bibitem{schmelzer2020machine}
M.~Schmelzer, R.~P. Dwight, P.~Cinnella, Discovery of algebraic reynolds-stress
  models using sparse symbolic regression, Flow, Turbulence and Combustion
  104~(2) (2020) 579--603.

\bibitem{wilcox2006turbulence}
D.~C. Wilcox, et~al., Turbulence modeling for CFD, Vol. 3rd Edition, DCW
  industries La Canada, CA, 2006.

\bibitem{hellsten2009explicit}
A.~Hellsten, S.~Wallin, Explicit algebraic reynolds stress and non-linear
  eddy-viscosity models, International Journal of Computational Fluid Dynamics
  23~(4) (2009) 349--361.

\bibitem{launder1975progress}
B.~E. Launder, G.~J. Reece, W.~Rodi, Progress in the development of a
  reynolds-stress turbulence closure, Journal of fluid mechanics 68~(3) (1975)
  537--566.

\bibitem{pope2001turbulent}
S.~B. Pope, Turbulent flows (2001).

\bibitem{versteeg2007introduction}
H.~K. Versteeg, W.~Malalasekera, An introduction to computational fluid
  dynamics: the finite volume method, Pearson education, 2007.

\bibitem{pope1975more}
S.~Pope, A more general effective-viscosity hypothesis, Journal of Fluid
  Mechanics 72~(2) (1975) 331--340.

\bibitem{Gatski1993}
T.~B. Gatski, C.~G. Speziale, {On explicit algebric stress models for complex
  turbulent flows}, Journal of Fluid Mechanics 254 (1993) 59--78.
\newblock \href {https://doi.org/10.1017/S0022112093002034}
  {\path{doi:10.1017/S0022112093002034}}.

\bibitem{XiaoCinnella2019}
H.~Xiao, P.~Cinnella, Quantification of model uncertainty in rans simulations:
  A review, Progress in Aerospace Sciences 108 (2019) 1--31.

\bibitem{duraisamy2019turbulence}
K.~Duraisamy, G.~Iaccarino, H.~Xiao, Turbulence modeling in the age of data,
  Annual Review of Fluid Mechanics 51, doi: 10.1146/annurev-fluid-010518-040547
  (2019).

\bibitem{parish2016paradigm}
E.~J. Paris, K.~Duraisamy, A paradigm for data-driven predictive modeling using
  field inversion and machine learning, Journal of Computational Physics 305
  (2016) 758.

\bibitem{ling2016reynolds}
J.~Ling, A.~Kurzawski, J.~Templeton, Reynolds averaged turbulence modelling
  using deep neural networks with embedded invariance, Journal of Fluid
  Mechanics 807 (2016) 155–166.
\newblock \href {https://doi.org/10.1017/jfm.2016.615}
  {\path{doi:10.1017/jfm.2016.615}}.

\bibitem{wu2018physics-informed}
J.~L. Wu, H.~Xiao, E.~Paterson, Physics-informed machine learning approach for
  augmenting turbulence models: A comprehensive framework, Physical Review
  Fluids 3 (2018) 074602.

\bibitem{weatheritt2016novel}
J.~Weatheritt, R.~Sandberg, A novel evolutionary algorithm applied to algebraic
  modifications of the rans stress--strain relationship, Journal of
  Computational Physics 325 (2016) 22--37.

\bibitem{weatheritt2017development}
J.~Weatheritt, R.~Sandberg, The development of algebraic stress models using a
  novel evolutionary algorithm, International Journal of Heat and Fluid Flow 68
  (2017) 298--318.

\bibitem{akolekar2019development}
H.~D. Akolekar, J.~Weatheritt, N.~Hutchins, R.~D. Sandberg, G.~Laskowski,
  V.~Michelassi, Development and use of machine-learnt algebraic reynolds
  stress models for enhanced prediction of wake mixing in low-pressure
  turbines, Journal of Turbomachinery 141~(4) (2019).

\bibitem{brunton2016discovering}
S.~L. Brunton, J.~L. Proctor, J.~N. Kutz, Discovering governing equations from
  data by sparse identification of nonlinear dynamical systems, Proceedings of
  the national academy of sciences 113~(15) (2016) 3932--3937.

\bibitem{rudy2017data}
S.~H. Rudy, S.~L. Brunton, J.~L. Proctor, J.~N. Kutz, Data-driven discovery of
  partial differential equations, Science Advances 3~(4) (2017) e1602614.

\bibitem{zhang2021customized}
Y.~Zhang, R.~Dwight, M.~Schmelzer, J.~Gómez, Z.-H. Han, S.~Hickel, Customized
  data-driven rans closures for bi-fidelity les–rans optimization, Journal of
  Computational Physics 432 (2021) 110153.
\newblock \href {https://doi.org/10.1016/j.jcp.2021.110153}
  {\path{doi:10.1016/j.jcp.2021.110153}}.

\bibitem{beetham2020formulating}
S.~Beetham, J.~Capecelatro, Formulating turbulence closures using sparse
  regression with embedded form invariance, Physical Review Fluids 5 (08 2020).
\newblock \href {https://doi.org/10.1103/PhysRevFluids.5.084611}
  {\path{doi:10.1103/PhysRevFluids.5.084611}}.

\bibitem{duraisamy2021perspectives}
K.~Duraisamy, Perspectives on machine learning-augmented reynolds-averaged and
  large eddy simulation models of turbulence, Physical Review Fluids 6 (2021)
  050504.

\bibitem{holland2019towards}
J.~R. Holland, J.~D. Baeder, K.~Duraisamy,
  \href{https://arc.aiaa.org/doi/pdf/10.2514/6.2019-1884}{Towards {Integrated}
  {Field} {Inversion} and {Machine} {Learning} {With} {Embedded} {Neural}
  {Networks} for {RANS} {Modeling}}, in: {AIAA} {Scitech} 2019 {Forum}, 2019,
  p. 1884.
\newline\urlprefix\url{https://arc.aiaa.org/doi/pdf/10.2514/6.2019-1884}

\bibitem{zhao2020rans}
Y.~Zhao, H.~D. Akolekar, J.~Weatheritt, V.~Michelassi, R.~D. Sandberg, Rans
  turbulence model development using cfd-driven machine learning, Journal of
  Computational Physics 411 (2020) 109413.

\bibitem{strofer2021end-to-end}
C.~A. Mich{\'e}len~Str{\:o}fer, H.~Xiao, End-to-end differentiable learning of
  turbulence models from indirect observations, Theoretical and applied
  mechanics letters 11 (2021) 100280.

\bibitem{banerjee2007presentation}
S.~Banerjee, R.~Krahl, F.~Durst, C.~Zenger, Presentation of anisotropy
  properties of turbulence, invariants versus eigenvalue approaches, Journal of
  Turbulence~(8) (2007) N32.

\bibitem{regis2005constrained}
R.~G. Regis, C.~A. Shoemaker, Constrained global optimization of expensive
  black box functions using radial basis functions, Journal of Global
  optimization 31~(1) (2005) 153--171.

\bibitem{Menter1994}
M.~F.R., {Two-equation eddy-viscosity turbulence model for engineering
  applications}, AIAA Journal 32 (1994) 1598--1605.

\bibitem{menter2012explicit}
F.~Menter, A.~Garbaruk, Y.~Egorov, Explicit algebraic reynolds stress models
  for anisotropic wall-bounded flows, Progress in Flight Physics 3 (2012)
  89--104.

\bibitem{weller1998tensorial}
H.~G. Weller, G.~Tabor, H.~Jasak, C.~Fureby, A tensorial approach to
  computational continuum mechanics using object-oriented techniques, Computers
  in physics 12~(6) (1998) 620--631.

\bibitem{caretto1973two}
L.~Caretto, A.~Gosman, S.~Patankar, D.~Spalding, Two calculation procedures for
  steady, three-dimensional flows with recirculation, in: Proceedings of the
  third international conference on numerical methods in fluid mechanics,
  Springer, 1973, pp. 60--68.

\bibitem{breuer2009flow}
M.~Breuer, N.~Peller, C.~Rapp, M.~Manhart, Flow over periodic hills--numerical
  and experimental study in a wide range of reynolds numbers, Computers \&
  Fluids 38~(2) (2009) 433--457.

\bibitem{marquillie2011instability}
M.~Marquillie, U.~Ehrenstein, J.-P. Laval, et~al., Instability of streaks in
  wall turbulence with adverse pressure gradient, Journal of Fluid Mechanics
  681~(205-240) (2011) 30.

\bibitem{bentaleb2012large}
Y.~Bentaleb, S.~Lardeau, M.~A. Leschziner, Large-eddy simulation of turbulent
  boundary layer separation from a rounded step, Journal of Turbulence~(13)
  (2012) N4.

\bibitem{brunton2019data}
S.~L. Brunton, J.~N. Kutz, Data-driven science and engineering: Machine
  learning, dynamical systems, and control, Cambridge University Press, 2019.

\bibitem{mcconaghy2011ffx}
T.~McConaghy, Ffx: Fast, scalable, deterministic symbolic regression
  technology, in: Genetic Programming Theory and Practice IX, Springer, 2011,
  pp. 235--260.

\bibitem{bishop2006pattern}
C.~M. Bishop, Pattern recognition and machine learning, springer, 2006.

\bibitem{knysh2016blackbox}
P.~Knysh, Y.~Korkolis, Blackbox: A procedure for parallel optimization of
  expensive black-box functions, arXiv preprint arXiv:1605.00998 (2016).

\bibitem{speziale1994realizability}
C.~G. Speziale, R.~Abid, P.~A. Durbin, On the realizability of reynolds stress
  turbulence closures, Journal of Scientific Computing 9~(4) (1994) 369--403.

\bibitem{emory2014visualizing}
M.~Emory, G.~Iaccarino, Visualizing turbulence anisotropy in the spatial domain
  with componentality contours, Center for Turbulence Research Annual Research
  Briefs (2014) 123--138.

\bibitem{edeling2018data}
W.~N. Edeling, G.~Iaccarino, P.~Cinnella, Data-free and data-driven rans
  predictions with quantified uncertainty, Flow, Turbulence and Combustion
  100~(3) (2018) 593--616.

\bibitem{kommenda2020parameter}
M.~Kommenda, B.~Burlacu, G.~Kronberger, M.~Affenzeller, Parameter
  identification for symbolic regression using nonlinear least squares, Genetic
  Programming and Evolvable Machines 21 (2020) 471--501.
\newblock \href {https://doi.org/https://doi.org/10.1007/s10710-019-09371-3}
  {\path{doi:https://doi.org/10.1007/s10710-019-09371-3}}.

\bibitem{pmlr-v48-yangc16}
Z.~Yang, Z.~Wang, H.~Liu, Y.~Eldar, T.~Zhang,
  \href{http://proceedings.mlr.press/v48/yangc16.html}{Sparse nonlinear
  regression: Parameter estimation under nonconvexity}, in: M.~F. Balcan, K.~Q.
  Weinberger (Eds.), Proceedings of The 33rd International Conference on
  Machine Learning, Vol.~48 of Proceedings of Machine Learning Research, PMLR,
  New York, New York, USA, 2016, pp. 2472--2481.
\newline\urlprefix\url{http://proceedings.mlr.press/v48/yangc16.html}

\bibitem{gloerfelt2019large}
X.~Gloerfelt, P.~Cinnella, Large eddy simulation requirements for the flow over
  periodic hills, Flow, Turbulence and Combustion 103~(1) (2019) 55--91.

\bibitem{wallin2000explicit}
S.~Wallin, A.~V. Johansson, An explicit algebraic reynolds stress model for
  incompressible and compressible turbulent flows, Journal of Fluid Mechanics
  403 (2000) 89--132.

\bibitem{durbin1991near}
P.~A. Durbin, Near-wall turbulence closure modeling without “damping
  functions”, Theoretical and computational fluid dynamics 3~(1) (1991)
  1--13.

\end{thebibliography}

\appendix

\section{The k-$\omega$ SST model}
\label{appendix:k-omega_SST}

Transport equations of $k$ and $\omega$ write :
\begin{equation}
\frac{\partial k}{\partial t} + U_j \frac{\partial k}{\partial x_j} = P_k -\beta ^ * k \omega + \frac{\partial }{\partial x_j} [(\nu + \sigma_k \nu_t) \frac{\partial k}{\partial x_j}]
\label{schmelzer_k_transport_baseline}
\end{equation}

\begin{equation}
\frac{\partial \omega}{\partial t} + U_j \frac{\partial \omega}{\partial x_j} =  \frac{\gamma}{\nu_t} P_k - \beta \omega^2 + \frac{\partial }{\partial x_j} [(\nu + \sigma_\omega \nu_t) \frac{\partial \omega}{\partial x_j}] + C D_{k \omega},
\label{schmelzer_omega_transport_baseline}
\end{equation} 

with $P_k=-2\nu_t S_{ij} \partial_j U_i$. The corresponding eddy viscosity writes

\begin{equation}
	\nu_t=\frac{a_1 k}{\textrm{max}(a_1 \omega , S F_2)}.
	\label{eq:k-omega-SST_nut}
\end{equation}

The other standard terms read

\begin{equation}
\begin{array}{l} 
C D_{k \omega} = \textrm{max} ( 2 \sigma_{\omega^2} \frac{1}{\omega} (\partial_i k)(\partial_i \omega) , 10^{-10} ) ,  \\
F_1 = \textrm{tanh} [( \textrm{min} [\textrm{max} (\frac{\sqrt{k}}{\beta^* \omega y } , \frac{500 \nu}{y^2 \omega}), \frac{4 \sigma_{\omega^2} k }{C D_{k \omega} y^2} ] )^4] , \\
F_2 = \textrm{tanh} [(\textrm{max} (\frac{2\sqrt{k}}{\beta^* \omega y } , \frac{500 \nu}{y^2 \omega}) ] )^2] , \\
\Phi = F_1 \Phi_1 + (1-F_1) \Phi_1,   
\end{array}	
\label{k_omega_basique_standard_terms}
\end{equation}
in which the latter blends the coefficients $\Phi \rightarrow (\Phi_1 , \Phi_2)$

\begin{equation}
\begin{array}{l l l l} 
\alpha = (5/9,0.44), & \beta=(3/40,0.0828), & \sigma_k=(0.85,1.0), & \sigma_\omega = (0.5,0.856).
\end{array}	
\label{k_omega_basique}
\end{equation}

The remaining terms are $\beta^*=0.09$, $a_1=0.31$.

\section{The BSL-EARSMmodel}
\label{appendix:BSL}

We here recall the formulation of the BSL-EARSM of Menter et al. \cite{menter2012explicit}. This model is based on the EARSM formulation of Wallin and Johansson \cite{wallin2000explicit} (WJ model) for the stress-strain relationship. 
In the WJ model, the stress-strain relationship is combined with the k-$\omega$ transport equations of Wilcox \cite{wilcox2006turbulence}. In the BSL-EARSM, in order to avoid the freestream sensitivity of the Wilcox model, the WJ stress-strain relationship is combined with the BSL k-$\omega$ model of Menter \cite{Menter1994}.

%
%

Following \cite{pope1975more}, the Reynolds stress anisotropy tensor $a_{ij}$ is projected onto a tensor basis :

\begin{equation}
a_{ij}=\beta_1 T_{1,ij} +\beta_2 T_{2,ij} +\beta_3 T_{3,ij} +\beta_4 T_{4,ij} +\beta_6 T_{6,ij} +\beta_9 T_{9,ij}      ,
\label{eq:aij_decomposition}
\end{equation}  
where
\begin{equation}
\begin{array}{l}
T_{1,ij}=S^*_{ij}; \quad T_{2,ij}=S^*_{ik}S^*_{kj}-\frac{1}{3} I_1 \delta_{ij}; \quad T_{3,ij}= \Omega^*_{ik}\Omega^*_{kj} -\frac{1}{3} I_2 \delta_{ij};\\
T_{4,ij}= S^*_{ik}\Omega^*_{kj}-\Omega^*_{ik}S_{kj}; \quad  T_{6,ij}= S^*_{ik} \Omega^*_{kl} \Omega^*_{lj} +  \Omega^*_{ik} \Omega^*_{kl} S_{lj} - \frac{2}{3} I_4 \delta_{ij} - I_2 S^*_{ij}; \\
T_{9,ij}=  \Omega^*_{ik} S^*_{kl} \Omega^*_{lm} \Omega^*_{mj} - \Omega^*_{ik} \Omega^*_{kl} S^*_{lm} \Omega^*_{mj} +\frac{1}{2} I_2 (S^*_{ik}\Omega^*_{kj} - \Omega^*_{ik} S^*_{kj})                 ,
\end{array}
\label{eq:tensorial_basis_BSL-EARSM}
\end{equation}
with $S^*_{ij}$ and $\Omega^*_{ij}$, the non-dimensional mean strain rate and rotation rate defined as follows :
\begin{equation}
S^*_{ij}=\frac{\tau}{2} (\frac{\partial U_i}{\partial x_j} + \frac{\partial U_j}{\partial x_i}), \quad \Omega^*_{ij}=\frac{\tau}{2} (\frac{\partial U_i}{\partial x_j} - \frac{\partial U_j}{\partial x_i}) 
\label{eq:S_{ij}_Omega_{ij}}
\end{equation}
where $\tau$ is a turbulent time scale with a Kolmogorov limiter \cite{durbin1991near}:
\begin{equation}
\tau=\text{max}(\frac{1}{C_\mu \omega},6 \sqrt{\frac{\nu}{C_\mu k \omega}}).
\label{eq:Kolmogorov_limiter}
\end{equation}
The tensor invariants $I_1$, $I_2$ and $I_4$ read:
\begin{equation}
I_1=S^*_{ij} S^*_{ji}, \quad I_2=\Omega^*_{ij} \Omega^*_{ji}, \quad I_4=S^*_{ik} \Omega^*_{kj} \Omega^*_{ji} .
\label{eq:invariants_BSL_EARSM}
\end{equation}
The coefficients of the tensor basis $\beta_i$ in \ref{eq:aij_decomposition} are defined as :

\begin{equation}
\beta_1= -\frac{N}{Q}, \quad \beta_2=0, \quad \beta_3=-\frac{2I_4}{NQ_1}, \quad \beta_4=-\frac{1}{Q}, \quad \beta_6=-\frac{N}{Q_1}, \quad \beta_9=\frac{1}{Q_1}, 	
\label{eq:coeffs_BSL-EARSM}	
\end{equation}

with

\begin{equation}
	Q=\frac{(N^2-2 I_2)}{A_1}, \quad Q_1=\frac{Q}{6}(2N^2-I_2)
	\label{eq:Q_Q1_BSL-EARSM}	
\end{equation}

where

\begin{equation}
N=C_1^\prime +\frac{9}{4} \frac{\tilde{P}_k}{\epsilon}
\label{eq:N_BSL-EARSM}	
\end{equation}

and

\begin{equation}
A_1=1.2, \quad C_1^\prime=\frac{9}{4}(C_1-1) \quad \text{and} \quad C_1=1.8.
\label{eq:A1_C1prime_C1_BSL-EARSM}	
\end{equation}

N is a solution of the cubic equation :

\begin{equation}
N^3 - C_1^\prime N^2 -(2.7 I_1 + 2 I_2) N + 2 C_1^\prime I_2 =0
\label{eq:N_equation_BSL-EARSM}	
\end{equation}

which is given by :

\begin{equation}
\left\{
\begin{array}{l}
    N= \frac{C_1^\prime}{3} + (P_1 + \sqrt{P_2})^{1/3} + \text{sign}(P_1-\sqrt{P2}) \mid P_1-\sqrt{P_2} \mid^{1/3} \quad \text{at} \quad P_2 \geq 0   \\
    N= \frac{C_1^\prime}{3} + 2 (P_1^2-P_2)^{1/6} \text{cos} (\frac{1}{3} \text{arccos}(\frac{P_1}{\sqrt{P_1^2-P_2}}))    \quad \text{at} \quad P_2 < 0      
\end{array}
\right.
\label{eq:N_solution_BSL-EARSM}
\end{equation}

with

\begin{equation}
P_1=C_1^\prime (\frac{C_1^{\prime 2}}{27} + \frac{9}{20} I_1 - \frac{2}{3} I_2), \quad P_2=P_1^2 - ( \frac{C_1^{\prime 2}}{9} + \frac{9}{10} I_1 +\frac{2}{3} I_2 )^3.
\label{eq:P1_P2_BSL-EARSM}	
\end{equation}

The BSL-EARSM constitutive equation is then supplemented by transport equations for k and $\omega$. These are the same as in \ref{appendix:k-omega_SST}, where a modified production term is used:
\begin{equation}
\tilde{P}_k = \text{min} (-\tau_{ij} \frac{\partial U_i}{\partial x_j}, 10.\rho \beta^* k \omega).
\label{eq:production_BSL-EARSM}	
\end{equation}
The reader is referred to \cite{menter2012explicit} for further details.

\section{CORS algorithm}\label{appendix:CORS}
The main steps of the algorithm proceed as follows:
\begin{itemize}
	\item The total number of function evaluations, noted N, is set from the beginning. 
%
%
	\item Initial step : 
	  \begin{itemize} 
	  	\item An initial sampling of size $n$ of the parameter space $S_1=\left\{\hat{\Theta}_i |i=1,2,\dots,n\right\}$ is performed using the Latin Hypercube sampling method. It must satisfy the condition $D<n<N$, where $D$ is the dimensionality of the search space.
	  	
	  	\item An initial response surface $\hat{f}_1$ is constructed using cubic radial basis functions (RBF) and the inititial sampling $S_1$. The functions are of the form:
	  	
	  	\begin{equation}
	  		\hat{f}_1(\hat{\Theta})=\sum_{i=1}^{N/2} \lambda_i \phi (\| \hat{\Theta} - \hat{\Theta}_i   \|) + b^T \hat{\Theta} +a,
	  	\label{eq:response_surface_1}	
	  	\end{equation}
	    where $\phi$ is a cubic function ($\phi(r)=r^3$) and $\lambda_i$, $b$ and $a$ are coefficients determined by interpolating the available samples.
	  \end{itemize}
  \item Steps 1 to $m=N-n$:
   \begin{itemize}
   	\item The candidate sample $\hat{\Theta}_i$ minimizing the surface response $\hat{f}_{i-1}$ is selected. In order to prevents the algorithm from being trapped in a local minimum, a ball of radius $r$ is placed around each of previously sampled points and the candidate sample minimizing the surface response is required to be outside of any ball. The radius of the balls $r$ decrease (all balls have the same radius) with iterations, with a given rate. The size of the balls is controlled by two parameters : their initial density $\rho_0$, and the rate of decay of their radius $p$. The density $\rho$ is the total volume of the balls divided by the total volume of the search space. At a given iteration $i$ ($1 \leq i \leq N/2$), the density and the radius are:
       \begin{equation}
           \rho_i=\rho_0 (\frac{m-i}{m-1})^p \quad ,
       \end{equation}  
       
       \begin{equation}
           r_i=(\frac{\rho_i}{(n+i-1)v_1})^{\frac{1}{D}},
       \end{equation}
       where $n$ and $m$ are respectively the number of initial samples and the number of subsequent steps; $D$ is the dimensionality of the search space and $v_1$ is a volume of a ball with radius 1, that can be expressed with the gamma function $\Gamma$ :
       \begin{equation}
           v_1=\frac{\pi^{\frac{D}{2}}}{\Gamma(\frac{D}{2}+1)} \quad .
       \end{equation}
   	\item The cost function is evaluated for $\hat{\Theta}_i$.
   	\item A new response surface $\hat{f}_{i}$ is constructed using the sampling $S_{i-1} \bigcup \hat{\Theta}_i$
   \end{itemize}
   
The candidate sample $\hat{\Theta}$ corresponding to the lowest evaluated value of the cost function is the solution of the problem proposed by the algorithm.

During both the initial and subsequent steps, RANS simulations are performed using candidate models in order to evaluate the cost function of the optimization problem. To force the research towards robust models, each model preventing the convergence of the solution is eliminated and a new model is resampled by introducing a slight perturbation into the non robust model. This procedure is repeated until a candidate model allowing convergence of the RANS simulation is obtained. 
\end{itemize}

\end{document}